\documentclass[preprint]{aastex}

\begin{document}

\title{The Statistical Discrepancy between the IGM and Dark Matter Fields:
  One-Point Statistics}

\author{Jes\'us Pando\altaffilmark{1}, Long-Long Feng\altaffilmark{2,3} and
                  Li-Zhi Fang\altaffilmark{4}}

\altaffiltext{1}{Department of Physics, DePaul University, Chicago
Il, 60614} \altaffiltext{2}{Purple Mountain Observatory, Chinese
Academy of Sciences, Nanjing, 210008} \altaffiltext{3}{National
Astronomical Observatories, Chinese Academy of Science, Chao-Yang
District, Beijing, 100012, P.R. China} \altaffiltext{4}{Department
of Physics, University of Arizona, Tucson, AZ 85721}

\begin{abstract}

We investigate the relationship between the mass and velocity fields of
the intergalactic medium (IGM) and dark matter. Although the evolution
of the IGM is dynamically governed by the gravity of the underlying
dark matter field, some statistical properties of the IGM inevitably
decouple from those of the dark matter once the nonlinearity of
the dynamical equations and the stochastic nature of the field
is considered. With simulation samples produced by a hybrid cosmological
hydrodynamic/N-body code, which is effective in capturing shocks and
complicated structures with high precision, we find that the one-point
distributions of the IGM field are systematically different from that
of dark matter as follows: 1.) the one-point distribution of the IGM
peculiar velocity field is exponential at least at redshifts less than 2,
while the dark matter velocity field is close to a  Gaussian field; 2.)
although the one-point distributions of the IGM and dark matter are 
similar, the point-by-point correlation between the IGM and dark matter 
density fields significantly differs on all scales and redshifts analyzed;
3.) the one-point density distributions of the difference between IGM 
and dark matter fields are highly non-Gaussian and long tailed. These 
discrepancies violate the similarity between the IGM and dark matter and 
cannot be explained simply as Jeans smoothing of the IGM. 
However, these statistical discrepancies are consistent with the 
fluids described by stochastic-force driven nonlinear dynamics.

\end{abstract}


\section{Introduction}

The mass field of the universe is dominated by dark matter with only
a tiny fraction of cosmic matter in the form of baryonic particles.
Gravitational clustering of dark matter is the major imputes for cosmic
structure formation. Collapsed dark matter halos host
various baryonic light-emitting and absorbing objects and information about
dark matter is available only via  baryons. The major component of baryonic
matter is in the form of gas, the intergalactic medium (IGM) (Here we use
IGM to mean baryonic gas both before and after the galaxy formation).
Therefore, the dynamical relationship between baryonic gas and the underlying
dark matter field is essential in order to understand the origin and
evolution of cosmic structures.

Since the IGM is only a tiny fraction of cosmic matter, it is usually
assumed that its clustering behavior on scales larger than the Jeans
length traces the underlying dark matter field. The density and velocity
distribution of the IGM is considered to be the same as the dark matter
field point-by-point on scales larger than the Jeans length scales.
That is
\begin{equation}
\delta_{igm}({\bf x}, t)= \delta_{dm}({\bf x}, t), \hspace{5mm}
{\bf v}_{igm}({\bf x}, t)= {\bf v}_{dm}({\bf x}, t)
\end{equation}
where $\delta_{igm}$ and $\delta_{dm}$ are the mass density contrasts
and ${\bf v}_{igm}$ and ${\bf v}_{dm}$ are the velocity fields, smoothed
on the Jeans length scales. Eq.(1) applies during the linear regime. Even
if the IGM is initially distributed differently than the dark matter,
linear growth modes will lead to eq.(1) on scales larger than the Jeans
length (Bi, B\"orner \& Chu 1992, Fang et al. 1993; Nusser 2000, Nusser
\& Haehnelt 1999, see also Appendix B). In this case, all statistical 
properties of the IGM at 
scales greater than the Jean's length are completely
determined by the dark matter field. In other words, the dynamical
behavior of the IGM field can be obtained from the dark matter field via a
similarity mapping (e.g., Kaiser 1986).

Because the nonlinear evolution of the IGM is also driven by the gravity
of the underlying dark matter, it is often assumed
that the relation (1) will hold even in the nonlinear regime. However,
this assumption is probably not valid. It has been shown in hydrodynamic
studies that a passive substance generally decouples from the underlying 
field
during nonlinear evolution (for a review, Shraiman \& Siggia 2001).
For instance, a passive substance might be highly non-Gaussian even when
the underlying field is Gaussian (Kraichnan 1994). This non-linear 
decoupling is
generic to systems consisting of a ``passive substance" and an underlying
stochastic mass field.

The IGM and dark matter fields interact stochastically since the initial
perturbations of the cosmic mass and velocity fields are randomly
distributed.  Their mass and velocity fields act as
random variables. The importance of the stochastic nature for the
evolution of cosmic mass and velocity field has been emphasized in
many studies (e.g. Berera \& Fang 1994, Jones, 1999, Buchert, Dominguez \&
Peres-Mercader 1999, Coles \& Spencer 2003, Ma \& Bertschinger 2003).
In this paper, we will address the statistical discrepancy between
the random fields of the IGM and dark matter in nonlinear regime.

Statistical discrepancies between the IGM and dark matter mass fields may
have already been detected in previous studies.  The adhesion model of the 
IGM shows that the deterministic relation 
$\delta_b({\bf x},t) = \delta({\bf x},t)$
no longer holds (Jones, 1999). The stochastic nature of cosmic mass field
indicates that the IGM should be described by a random-force driven Burgers'
equation (Matarrese \& Mohayee 2002) which does not always yield the 
deterministic solution 
$\delta_b({\bf x},t) = \delta({\bf x},t)$. Cosmological hydrodynamic
simulations also find that $\rho_{igm}$ does not tightly correlate with
$\rho_{dm}$, but is largely scattered around the line
$\delta_{igm}=\delta_{dm}$, with this scatter not due to
noise (Gnedin \& Hui 1998). It has also been found that the scatter defined 
by
\begin{eqnarray}
\bigtriangleup \delta({\bf x},t)
&=&\delta_{dm}({\bf x}, t) - \delta_{igm}({\bf x}, t)  \\ \nonumber 
\bigtriangleup {\bf v}({\bf x},t)
&=&{\bf v}_{dm}({\bf x}, t) - {\bf v}_{igm}({\bf x}, t)
\end{eqnarray}
is highly non-Gaussian (Feng, Pando \& Fang 2003). This strengthens the
conclusion that the discrepancy between $\delta_{igm}$ and $\delta_{dm}$ is
not due to computational noise or other Gaussian processes,
but probably arises from the nonlinear evolution of the random fields.

Here we study this discrepancy at a more fundamental level. As a first step,
we concentrate on the density and velocity field one-point IGM distributions,
and their discrepancy from dark matter. The outline of this paper is as
follows. \S 2 addresses the dynamical mechanisms, and gives predictions on the
statistical discrepancy between the IGM and dark matter density and velocity
fields. \S 3 presents the cosmological hydro simulation scheme and samples.
The relevant one-point statistics are developed in \S 4. \S 5 shows the
discrepancy and tests the predictions with one-point statistics of the density
and velocity fields. Finally, conclusions and discussions are in \S 6.

\section{Statistical discrepancy between the IGM and dark matter fields }

During the linear regime, thermal diffusion leads to a discrepancy between the
IGM field and dark matter field on scales up to the
Jeans length. That is, the IGM mass density perturbations
are suppressed on scales less than the Jeans length. The IGM
density perturbations on scales larger than the Jeans length are Gaussian if
the dark matter density perturbations are Gaussian.
However, in the nonlinear regime, a statistical discrepancy very different
from a simple thermal diffusion can appear.
This can be illustrated by considering the isothermal model of the IGM. In
this case, the IGM density is given by
$\rho_{igm}({\bf x}) \propto \exp[-m\phi({\bf x})/k_BT]$, where
$\phi({\bf x})$ is gravitational potential, $T$ local temperature, $m$
the mass of the IGM particles and $k_B$ the Boltzmann constant. If
$\phi({\bf x})$ is a Gaussian random field, $\rho_{igm}$ will
be a lognormal random field (Zeldovich, Ruzmaikin \& Sokoloff 1990) and
a statistical discrepancy between $\rho_{igm}({\bf x})$ and dark matter
arises.  Although the isothermal model is not realistic on large scales in
general, it reveals that the statistical discrepancy between the IGM and dark
matter might arise if 1.) the IGM evolution is nonlinear and 2.) the relevant
fields, like $\phi({\bf x})$, are stochastic.

\subsection{The statistical discrepancy of velocity fields}

A more realistic discrepancy appears when considering the differences
between the dynamics of the peculiar velocity fields of the IGM
${\bf v}_{igm}$ and dark matter ${\bf v}_{dm}$. Since dark matter particles are
collision-less, the intersection of the dark matter particle trajectories
will lead to a multi-valued velocity field. On the other hand, as a fluid,
the velocity field of the IGM will always be single-valued. At
the intersection of dark matter particle trajectories, the IGM velocity
field will be discontinuous and yield shocks or complicated structures
(Shandarin \& Zeldovich 1989). Shocks in the IGM can significantly
change the mass density and velocity of the baryon gas, but will exert no
direct effect on the dark matter field.  It is at this point that  the
dynamical similarity between the IGM and dark matter is broken.

We analyze this situation using the dynamical equations for dark matter
and the IGM. For growth modes, the peculiar velocity field
of the dark matter is vortex-free. One can define a velocity
potential $\varphi$ as ${\bf v}_{dm}=-(1/a)\nabla \varphi_{dm}$, where $a$
is the cosmic scale factor. The dynamical equation of the velocity potential is
(see Appendix \S A)
\begin{equation}
\frac{\partial \varphi_{dm}}{\partial t}-
\frac{1}{2a^2}(\nabla \varphi_{dm})^2 = \phi,
\end{equation}
where $\phi$ is the gravitational potential and depends on density perturbation
$\delta_{dm}=[\rho_{dm}-\overline{\rho}_{dm}]/\overline{\rho}_{dm}$ via 
the Poisson equation (A3). Generally, the field $\phi$ is Gaussian, or only 
slightly deviates from a
Gaussian field with $\langle \phi \rangle =0$ and variance
$\langle\phi^2\rangle$. Equation (3) is valid till
the intersection or shell crossing of dark matter particle trajectories
has occurred.

For the IGM growth modes, one can also define a velocity potential 
$\varphi_{igm}$, by ${\bf v}_{igm}= -(1/a)\nabla \varphi_{igm}$. The 
dynamical equation (3) are approximately given by (see Appendix \S B)
\begin{equation}
\frac{\partial \varphi_{igm}}{\partial t}-
\frac{1}{2a^2}(\nabla \varphi_{igm})^2 -
\frac{\nu}{a^2}\nabla^2 \varphi_{igm} =\phi.
\end{equation}
The coefficient $\nu$ is given by
\begin{equation}
\nu=\frac{\gamma k_BT_0}{\mu m_p (d \ln D(t)/dt)},
\end{equation}
where $D(t)$ describes the linear growth behavior. The diffusion term
$\nu$ in eq.(4) is given by the Jeans smoothing. The wavenumber of
the comoving Jeans scale is $k_J^2=(a^2/t^2)(\mu m_p/\gamma k_BT_0)$,
where $m_p$ is proton mass. The parameters $\gamma$ and $\mu$ are,
respectively, the polytropic index and molecular weight of the IGM.

The nonlinear equation (4) is actually the stochastic-force driven
Burgers' equation or the KPZ equation (Kardar, Parisi \& Zhang 1986;
Berera \& Fang 1994). Fields governed by eq.(5) have been extensively
studied to model structure formation (Barab\'asi \& Stanley  1995).
The solution to eq. (4) depends on two characteristic scales: 1.) the 
correlation length $r_c$ of the random field $\phi$; 2.) the dissipation 
length 
$d=\nu^{3/4}r_c^{1/2}\langle\phi^2\rangle ^{-1/4}(d\ln D/dt)^{-1/4}$,
which is due to Jeans smoothing. The behavior of the field 
$\varphi_{igm}$ governed by eq.(4) is determined by its Reynolds number 
defined as ${\cal R}\equiv (r_c/d)^{4/3}$ (e.g., L\"assig 2000) or
\begin{equation}
{\cal R} =(k_Jr_c)^{2/3}\left(\frac{k_J}{k}\right )^{4/3}
    \langle \delta_{dm}^2(k)\rangle^{1/3}
\end{equation}
where $r_c$ is the comoving correlation length, and $\delta_{dm}(k)$ is
the Fourier component of the density contrast on wavenumber $k$. To derive
eq.(6), we assume that the gravitational potential $\phi$ is only
given by the dark matter mass perturbation. When the Reynolds number
is larger than 1, Burgers turbulence occurs in the $\varphi_{igm}$ field.

The correlation length $r_c$ of the gravitational potential $\phi$ is
larger than the Jeans length and we have $k_Jr_c >1$. Therefore,
${\cal R}$ can be larger than 1 on scales larger than the Jeans length,
even when $\delta_{dm}(k_J)$ is on order 1. That is,
Burgers turbulence develops in the IGM $\varphi_{igm}$ field
while the dark matter mass density perturbations are still quasi-linear
or weakly nonlinear. We can conclude that
{\em the evolution of the PDFs of ${\bf v}_{igm}({\bf x},t)$ and 
${\bf v}_{dm}({\bf x},t)$ should be significantly different, the 
former becoming non-Gaussian earlier than the latter.}

When Burgers turbulence develops in the IGM, $\varphi_{igm}$ is
characterized by strong intermittency and contains
discontinuities, or shocks. The probability distribution function
(PDF) of $\varphi_{igm}$ is long tailed (L\"assig 2000). The
intermittent spikes are the events that populate the long tail.
This feature is supported by observations of the  Ly$\alpha$ flux
transmission (Jamkhedkar, Zhan \& Fang 2000, Pando et al. 2002, 
Jamkhedkar et al. 2003). That is,
the randomly distributed shocks and intermittent spikes lead to
the statistical discrepancy between the velocity fields of the IGM
and dark matter. One can then expect that {\em the PDF of the
difference, $\bigtriangleup{\bf v}({\bf x})$, is long tailed.}
Moreover, the non-trivial difference between the
IGM and dark matter holds on the scales larger than the Jeans
length of the IGM.

\subsection{The 2$^{nd}$ moment of the density distributions}

The non-trivial velocity difference $\bigtriangleup{\bf v}({\bf x},t)$
will lead to a non-trivial density difference between the IGM and dark 
matter. The linearized continuity equations for dark matter [eq.(A1)]
and IGM [eq.(B1)] are, respectively
\begin{eqnarray}
\frac{\partial \delta_{dm}}{\partial t} +
  \frac{1}{a}\nabla \cdot {\bf v}_{dm} & = & 0
 \\
\frac{\partial \delta_{igm}}{\partial t} +
  \frac{1}{a}\nabla \cdot {\bf v}_{igm} & = & 0.
\end{eqnarray}
Thus, using the definition for $\bigtriangleup \delta({\bf x}, t)$,
we have
\begin{equation}
\bigtriangleup \delta({\bf x})\equiv
  \delta_{dm}- \delta_{igm} = - \int dt\nabla \cdot
\bigtriangleup {\bf v}.
\end{equation}
This result shows that the IGM mass field
$\delta_{igm}({\bf x})$ does not trace the dark matter field
$\delta_{dm}({\bf x})$ point-by-point due to the discrepancy in
their velocity fields.

The discrepancy in the mass fields can be measured by the 2$^{nd}$
moments of the density fields defined as
\begin{equation}
I_{dm}=\frac{\langle[\bigtriangleup\delta({\bf x})]^2\rangle^{1/2}}
    {\langle\delta^2_{dm}({\bf x})\rangle^{1/2}},
\end{equation}
\begin{equation}
I_{igm}=\frac{\langle[\bigtriangleup\delta({\bf x})]^2\rangle^{1/2}}
    {\langle\delta^2_{igm}({\bf x})\rangle^{1/2}},
\end{equation}
and
\begin{equation}
I=\frac{\langle[\bigtriangleup\delta({\bf x})]^2\rangle^{1/2}}
    {\langle[\delta_{igm}({\bf x})+
       \delta_{dm}({\bf x})]^2\rangle^{1/2}}.
\end{equation}
If $\delta_{igm}({\bf x})$ is perfect tracer of $\delta_{dm}({\bf
x})$, we have $I_{dm}=I_{igm}=I=0$. If both $\delta_{igm}({\bf x})$ and
$\delta_{igm}({\bf x})$ are statistically independent, we have
$I=1$. Thus, from eq.(9), we expect that {\em $I$, $I_{dm}$ and
$I_{igm}$ are smaller at higher redshift, and larger at lower
redshifts. }

\subsection{The density PDF discrepancy}

 From the continuity equations eqs.(A1), eq.(B1) and the definitions
eq.(2), we have
\begin{equation}
\frac{\partial \bigtriangleup \delta({\bf x})}{\partial t}= -\frac{1}{a}
\nabla \cdot \left[ \bigtriangleup {\bf v} + 
  (\bigtriangleup \delta) {\bf v}_{dm}
+ (\bigtriangleup {\bf v}) \delta_{igm}\right].
\end{equation}
We define a window sampling given by
\begin{equation}
\bigtriangleup^R({\bf x_0}) =\int W_R({\bf x'-x}_0)
   \bigtriangleup \delta({\bf x'})d{\bf x'},
\end{equation}
where the normalized window function $W_R({\bf x'-x}_0)$ is non-zero
around ${\bf x_0}$ on spatial scale $R$. Since $\bigtriangleup \delta$
and $\bigtriangleup \delta_{igm}$ are statistically isotropic, for an 
isotropic window function $W_R({\bf x'})$  the
terms containing $ \nabla\bigtriangleup \delta$ and $ \nabla\delta_{igm}$ 
are negligible. Thus, eq.(13) becomes a Langevin equation for 
$\bigtriangleup^R({\bf x})$ 
\begin{equation}
\frac{d \bigtriangleup^R}{d t}= -
\frac{1}{a} \lambda\bigtriangleup^R + \frac{1}{a}\eta,
\end{equation}
The first term on the r.h.s. of eq.(15) is a friction term with
friction coefficient given by
\begin{equation}
\lambda = g_W\int W_R({\bf x'-x}_0)
   (\nabla_{\bf x'} \cdot {\bf v}_{dm}) d{\bf x'},
\end{equation}
where the factor $g_W$ depends only on the window function (see \S 5.3).
The random driving force $\eta$ of eq.(15) is
\begin{equation}
\eta =- \int d{\bf x'} W_R({\bf x'-x}_0) \,
(\nabla \cdot \bigtriangleup {\bf v})(1+\delta_{igm}).
\end{equation}
Since $|\delta_{igm}| < 1 $ at most places, we can drop $\delta_{igm}$ 
in eq.(17).

Eq.(15) contains an additive stochastic force, $\eta$, and a
multiplicative stochastic force, $\lambda$. Both stochastic forces are 
given by the random velocity field. The Langevin equation
eq.(15) with both additive and multiplicative noise is typically used to
model intermittent fields (Graham, H\"ohnerbach \& Schenzle, 1982; Platt,
Hammel \& Heagy 1994, Nakao 1998). In the nonlinear regime, most locations in 
the dark matter density field have $\delta_{dm}<0$, and therefore,
$\nabla \cdot {\bf v}_{dm}$ and $\lambda$ are larger than zero. The
friction term $\lambda$ leads to the decay of $\bigtriangleup^R$. That is,
the density perturbation difference $\bigtriangleup^R$ caused by the
noise $\eta$ tends to zero on average. Although $\lambda$ is positive in
most cases, it can go negative due to fluctuations. In these cases,
$\bigtriangleup^R$ will be amplified exponentially and attain large
values, which are the spikes in the field. This is intermittency and
the PDF of $ \bigtriangleup^R$ will be long-tailed. It has been
shown that when the stochastic terms $\eta$ and $\lambda$ are Gaussian,
the PDF of $\bigtriangleup^R$ is a power law in general (Appendix C).
Although $\eta$ and $\lambda$ for the dark matter given by eqs.(16) and
(17) are not Gaussian, we still can conclude that the PDF of
$\bigtriangleup^R$ is generally long tailed, as long tails are a common
result of a multiplicative stochastic force. For instance, if one ignores the
additive noise term $\eta$, the solution for eq.(15) is
$\bigtriangleup^R \propto \exp(-\int \frac{1}{a} \lambda dt)$. Thus, the PDF 
tail $\bigtriangleup^R$ is longer than that of $\lambda$ so that,
for instance,  when $\lambda$ is Gaussian, $\bigtriangleup^R$ is lognormal.
Using Appendix eq.(C7), we see that
{\em the PDF of $\bigtriangleup^R$ is generally highly non-Gaussian on
   scales larger than the Jeans length. It is flat in the central part
  and gradually becomes a power law with index not lower than -1.}

By closely studying the dynamics of the velocity fields we have found that
the statistical discrepancy between the dark matter and IGM velocity
fields will manifest itself as follows:

\begin{enumerate}
\item The evolution of the PDF's of ${\bf v}_{igm}$ and ${\bf v}_{dm}$ 
will be significantly different. The former becomes non-Gaussian earlier 
than the latter. Further the PDF of the difference $\bigtriangleup${\bf v(x)}
will be long tailed.
\item The second moments $I,I_{dm},$ and $I_{igm}$ (eqs 10-12) will be
smaller at higher redshifts and larger at smaller redshifts.
\item The PDF of $\bigtriangleup^R$ are generally highly non-Gaussian on
   scales larger than the Jeans length.
\end{enumerate}
In the following sections we will test these predictions

\section{Hydrodynamic simulations}

\subsection{WENO hydrodynamic simulations}

To simulate the IGM, we use the the hydrodynamic equations
of the IGM in the form of conservation laws, eqs.(B4)-(B6). Although the
momentum equation (B5) is the typical Navier-Stokes equation,
gravitational instability leads to a system dominated by growth
modes and the dynamical equations essentially become a Burgers equation if 
only the growth modes are considered (Berera \& Fang 1994).
It is well known that the Burgers equation does not reduce initial chaos,
but increases it (Kraichman 1968). That is, when the Reynolds number is high,
an initially random field always yields a collection of shocks with a
smooth and simple variation of the field between the shocks. We conclude
that an optimal simulation scheme should capture shock and discontinuity
transitions as well as to calculate piecewise smooth functions with a
high resolution.

For these reasons we do not use schemes based on smoothed particle
hydrodynamic (SPH) algorithms. It is well known that one of the main
challenges to the SPH scheme is how to handle shocks or discontinuities
because SPH schemes smooth the fields. This
problem is not yet well settled (e.g. Borve, Omang, \& Trulsen,
2001, Omang, Borve, \& Trulsen 2003). Instead, we will take an Eulerian 
approach to simulating the IGM. However
there is a basic problem in Eulerian based codes in that they cause 
unphysical
oscillations near a  discontinuity. An effective method to reduce the
spurious oscillations is given by designed limiters, such as the
total-variation diminishing (TVD) schemes (Harten, 1983). However,
TVD accuracy degenerates to first order near
smooth extrema (Godlewski \& Raviart 1996). This problem is serious
in calculating the difference between fluid quantities on the two
sides of the shock when the Mach number of a gas is high. But this is
exactly the case in the gravitationally coupled IGM and dark matter
system. In this system, the IGM temperature is generally around
$10^{4-6}$ K and the sound speed a few km s$^{-1}$ to a few tens km
s$^{-1}$, while the IGM rms bulk velocity is on order of hundreds km
s$^{-1}$ (Zhan \& Fang 2002). Hence, the Mach number of the IGM can be
as high as $\sim$100 (Ryu et al 2003).

To overcome this problem, two algorithms ENO and then WENO were 
developed (Harten et al. 1987, Shu 1998, Fedkiw,
Guillermo \& Shu 2003, Shu, 2003.) The precision of the WENO scheme is
found to be higher than that of TVD and the piecewise parabolic method
(PPM) for both strong and weak shocks. TVD schemes degenerate to first-order
accuracy at locations of smooth extrema, while the ENO and WENO schemes
maintain their high-order accuracy. Both ENO and WENO use the idea of
adaptive stencils in the reconstruction procedure based on the local
smoothness of the  numerical solution to automatically achieve high
order accuracy and non-oscillatory properties near discontinuities (Liu,
Osher, \& Chan 1994; Jiang \& Shu 1996). This scheme is probably the
first successful attempt to obtain a self-similar (no mesh size dependent
parameter), uniformly high order accurate, yet essentially
non-oscillatory, interpolation for piecewise smooth functions.
WENO is robust and stable and simultaneously provides high order
precision for both the smooth part of the solution and sharp shock
transitions.

WENO has been successfully applied to hydrodynamic problems
containing shocks and complex structures, such as shock-vortex
interaction (Grasso \& Pirozzoli, 2000a, 2000b), interacting blast
waves (Liang \& Chen, 1999; Balsara \& Shu 2000), Rayleigh-Taylor
instability (Shi, Zhang \& Shu 2003), and magneto-hydrodynamics
(Jiang \& Wu, 1999). WENO has also been used to study
astrophysical hydrodynamics, including stellar atmospheres (Zanaa,
Velli \& Londrillo, 1998), high Reynolds number compressible flows
with supernova (Zhang et al. 2003), and high Mach number
astrophysical jets (Carrillo et al. 2003). In the context of
cosmological applications, WENO has proved especially adept at
handling the Burgers' equation (Shu 1999). Recently, a hybrid
hydrodynamic/N-body code based on the WENO scheme was developed
and passed typical reliability tests including the Sedov blast
wave and the formation of the Zeldovich pancakes (Feng, Shu \&
Zhang 2004). This code has been successful in producing the QSO
$Ly\alpha$ transmitted flux, including the high resolution sample
HS1700+6416 (Feng, Pando \& Fang 2003). The statistical features
of these samples are in good agreement with observed features not
only on second order measures, like the power spectrum, but also
to orders as high as eighth order for the intermittent behavior.
The code also has been shown to be effective in capturing gravitational 
shocks during the large scale structure formation (He, Feng and 
Fang 2004). Hence we believe that for the purposes of this work the 
Eulerian code based on the WENO scheme is the best approach.

\subsection{Samples}

For the present application, we run the hybrid N-body/hydrodynamic code to
trace the cosmic evolution of the coupled system of dark matter and
baryonic gas in a flat low density CDM model ($\Lambda$CDM), which is
specified  by the cosmological parameters
$(\Omega_m,\Omega_{\Lambda},h,\sigma_8,\Omega_b)=(0.3,0.7,0.7,0.9,0.026)$.
The baryon fraction is fixed with the constraint from primordial
nucleosynthesis as $\Omega_b=0.0125h^{-2}$ (Walker et al. 1991). The
linear power spectrum is taken from the fitting formulae presented by
Eisenstein \& Hu (1998).

Atomic processes including ionization, radiative cooling and heating are
modeled similarly as in Cen (1992) in a plasma of hydrogen and helium of
primordial composition ($X=0.76$, $Y=0.24$). Processes such as star
formation, and feedback due to SN and AGN activities are not taken into
account
as yet. A uniform UV-background of ionizing photons is assumed to
have a power-law spectrum of the form $J(\nu) =J_{21}\times10^{-21}
(\nu/\nu_{HI})^{-\alpha}$erg s$^{-1}$cm$^{-2}$sr$^{-1}$Hz$^{-1}$,
where the photo-ionizing flux is normalized by the parameter
$J_{21}$ at the Lyman limit frequency $\nu_{HI}$, and is suddenly
switched on at $z > 10$ to heat the gas and re-ionize the universe.

The simulations are performed in a periodic, cubic box of size
25 h$^{-1}$Mpc with a 192$^3$ grid and an equal number of dark
matter particles. The simulations start at a redshift
$z=49$ and the results are output at redshifts $z=$4.0, 3.0, 2.0,
1.0, 0.5 and 0.0. At each output stage, we read $\rho_{dm}$, $\rho_{igm}$
(both are, respectively, in the units of $\overline{\rho}_{dm}$
and $\overline{\rho}_{igm}$), ${\bf v}_{dm}$ and ${\bf v}_{igm}$
in each cell.

The time step is chosen by the minimum value among
the following three time scales. The first is from the Courant
condition given by
\begin{equation}
 \delta t \le \frac{ cfl \times a(t) \Delta x}{\hbox{max}(|v_x|+c_s,
|v_y|+c_s, |v_z|+c_s)}
\end{equation}
where $\Delta x$ is the cell size, $c_s$ is the local sound speed, $v_x$,
$v_y$ and $v_z$ are the local fluid velocities and $cfl$ is the Courant
number, which we take as $cfl=0.6$. The second time scale is imposed by
cosmic expansion which requires that $\Delta a /a <0.02$ within a single
time step. The last time scale comes from the requirement that a particle
moves not more than a fixed fraction of the cell size.

The effect of the numerical resolution of these samples has been
tested to higher order statistics(Feng, Pando \& Fang, 2003). It was
found that
two samples with different resolutions give about the same statistical
result in the intermittency out to order eight.

For statistical studies, we randomly sample 500 one-dimensional (1D) fields
from the simulation results at redshifts z=4, 3, 2, 1, 0.5 and 0.
Each 1D sample, covering size $L$=25 h$^{-1}$Mpc, contains
192 data points, each one containing the mass density and peculiar
velocity of dark matter, and the mass density and peculiar velocity of 
the IGM.

\subsection{$\rho_{igm}$ vs. $\rho_{dm}$}

\begin{figure}[h]
\plotone{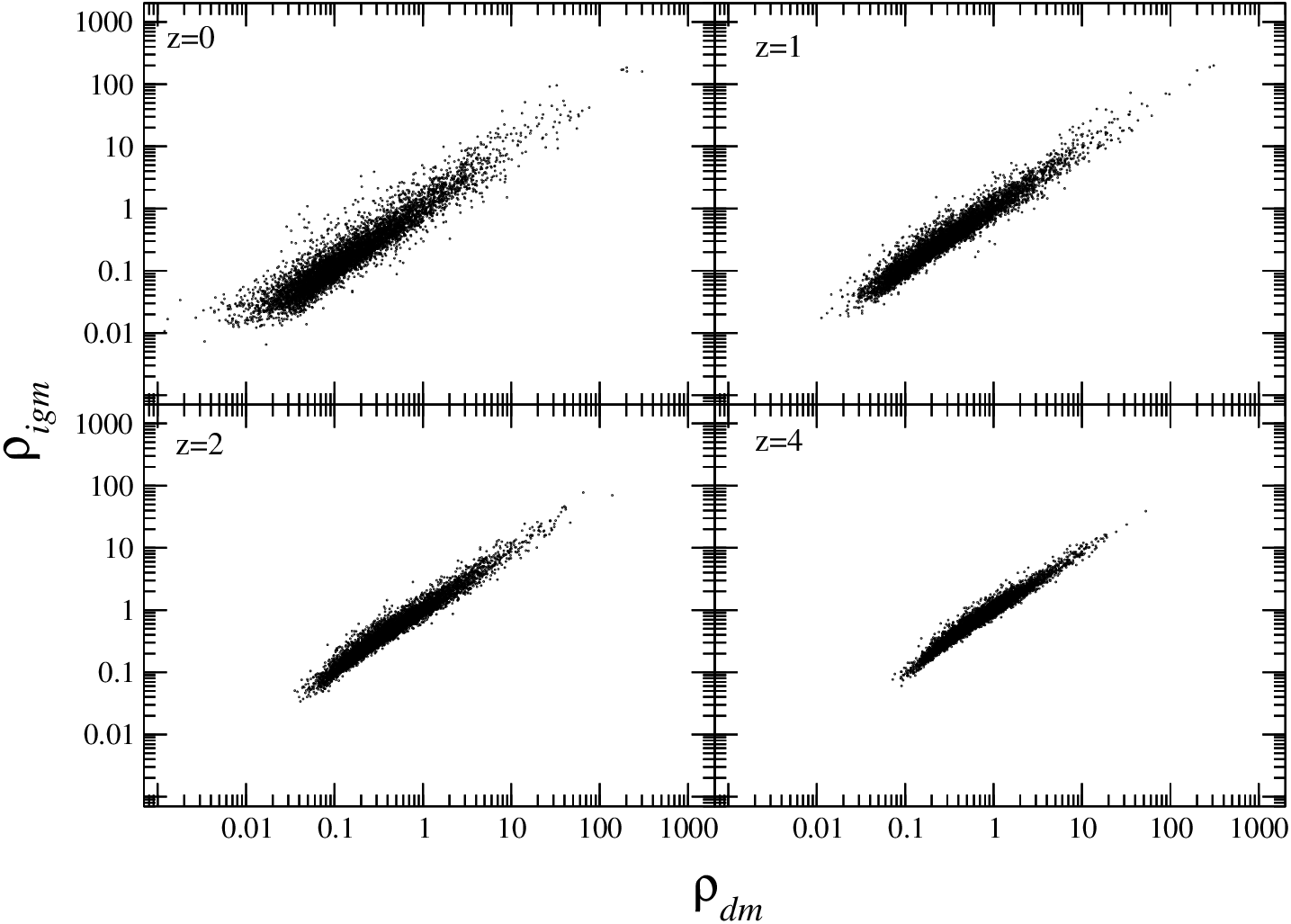}
\caption{he relation between $\rho_{igm}(x)$ and $\rho_{dm}(x)$ for
redshifts $z=0$, 1, 2 and 4. The data consists of $\simeq$ 7000 randomly
drawn points from the simulation sample }
\end{figure}

As first evidence of the discrepancy between the IGM and dark matter,
Fig. 1 gives the relation between $\rho_{dm}$ and $\rho_{igm}$
at redshifts 4, 2, 0.5 and 0. If $\rho_{dm}$ traces $\rho_{igm}$
point-by-point, we should have a tight correlation along the line
$\rho_{dm} = \rho_{igm}$. However, Fig. 1 shows that although the relation
$\rho_{dm} = \rho_{igm}$ is correct on average, the data points are
significantly scattered around the line $\rho_{dm} = \rho_{igm}$ in the 
entire density range from $\rho_{dm}\sim$ 0.01 to $10^2$.

The scatter of the IGM vs. dark matter has been noted in SPH
simulations (Gnedin \& Hui 1998). They found that the IGM density contrast
is significantly different from that of dark matter. They correctly pointed
out that a pure N-body simulation would fail to reproduce the IGM distribution
without mimicking the dynamical effect of the gas. They explain the scatter
as a dynamical effect of gas pressure. The pressure of the IGM with
temperature $\sim 10^4$ K is higher at higher redshifts, as the mean
density of the gas is higher at earlier times. However, Fig. 1 shows that
the scatter is smaller at higher redshifts. This would seem to contradict
the gas pressure as an explanation for the scatter.  On the other hand, the 
higher the nonlinearity, the higher the Reynolds number [eq.(6)], and the 
higher the discrepancy. Therefore, the gravitational nonlinear evolution of 
the IGM and dark matter system provides a plausible explanation of the 
scatter of Fig. 1.

\section{One-point Statistics}

\subsection{One-point variables with the DWT decomposition}

To calculate the one point distribution, we use the discrete
wavelet transform (DWT). The scaling functions of the DWT analysis
serve as the sampling window function. For the details of the
mathematical properties of the DWT see Mallat (1989a,b);
Meyer (1992); Daubechies (1992), and for cosmological applications see
Fang \& Thews (1998), Fang \& Feng (2000).

Let us briefly introduce the DWT-decomposition for a random field.
Consider a 1-D density fluctuation $\delta(x)$ on a spatial range
from $x=0$ to $L$. We divide the space into $2^j$ segments
labeled by $l=0,1,...2^j-1$ each of size $L/2^j$. The index
$j$ is a positive integer and gives the length scale $L/2^j$.
The larger the $j$ is, the smaller the length scale. Any reference
to a property as a function of scale $j$ below must be interpreted as
the property at length scale $L/2^j$. The index $l$ represents
position and it corresponds to the spatial range
$lL/2^j < x < (l+1)L/2^j$.  Hence, the space $L$ is decomposed into
cells $(j,l)$.

The discrete wavelet is constructed such that each cell $(j,l)$ supports a
compact function, the scaling function $\phi_{j,l}(x)$, which satisfies the
orthonormal relation
\begin{equation}
\int \phi_{j,l}(x)\phi_{j,l'}(x)dx = \delta^K_{l,l'},
\end{equation}
where $\delta^K$ is Kronecker delta function.
The scaling function $\phi_{j,l}(x)$ is a window function
on scale $j$ centered around the segment $l$.

For a field $F(x)$, its mean in cell $(j,l)$ can be estimated by
\begin{equation}
F_{j,l}=\frac{\int_{0}^{L} F(x)\phi_{j,l}(x)dx}
   {\int_{0}^{L} \phi_{j,l}(x)dx}=
   \frac {1}{\int_{0}^{L} \phi_{j,l}(x)dx} \epsilon^{F}_{j,l},
\end{equation}
where $\epsilon^{F}_{j,l}$ is called scaling function coefficient (SFC),
given by
\begin{equation}
\epsilon^{F}_{j,l}= \int_{0}^{L} F(x)\phi_{j,l}(x)dx.
\end{equation}

Thus, a 1-D field $F(x)$ can be decomposed into
\begin{equation}
F(x) =
  \sum_{l=0}^{2^j-1}\epsilon^{F}_{j,l}\phi_{j,l}(x) + O(\geq j).
\end{equation}
The term $O(\geq j)$ in eq.(22) contains only the
fluctuations of the field $F(x)$ on scales equal to and less
than $L/2^j$. This term does not have any contribution to
the window sampling on scale $j$. Thus, for a given $j$, the one-point
variables $F_{j,l}$ or $\epsilon^F_{j,l}$ ($l=0, 1...2^j-1$)
give a complete description of the field $F(x)$ smoothed on scale
$L/2^j$. As one-point variables the
$\epsilon^{F}_{j,l}$ are similar to the measure given by
count-in-cell technique. However, the orthonormality eq.(19) insures that
the set of $F_{j,l}$ or $\epsilon^F_{j,l}$ do not cause false
correlations.  In the calculations below, we use the
Daubechies 4 (D4) wavelet (Daubechies, 1992).

\subsection{One-point variables of density fields}

\begin{figure}[h]
\plotone{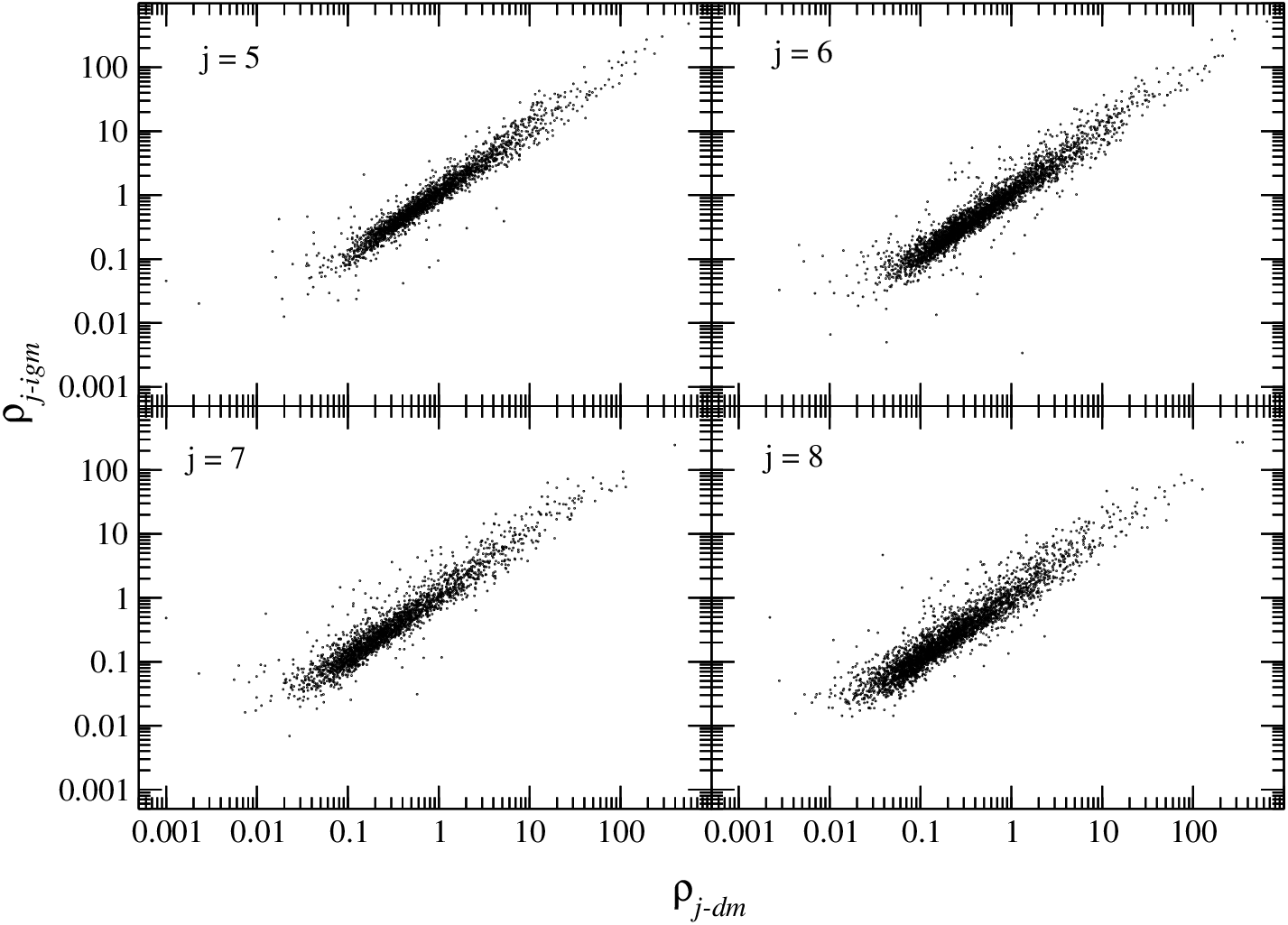}
\caption{The relation between $\rho_{j,l -igm}$ and $\rho_{j,l -dm}$ on
scales $j=5$, 6, 7 and 8, at redshifts $z=0$. The
comoving scale corresponding to $j$ is $33/2^{j}$ h Mpc$^{-1}$. The data
consists of $\simeq$ 7000 randomly drawn points from the simulation sample}
\end{figure}

We first calculate the one-point distributions of variables
$\rho_{j,l - igm}$ and $\rho_{j,l - dm}$, which are the
one-point variables given by eq.(20) replacing $F(x)$ by the density
fields $\rho_{igm}(x)$ and $\rho_{dm}(x)$. We then plot in Fig. 2
$\rho_{j,l - igm}$ vs.  $\rho_{j,l - dm}$ for $j=5$, 6, 7 and 8,
corresponding to comoving length scales 1.03, 0.516, 0.258 and
0.129 h$^{-1}$ Mpc, respectively. Fig. 2 shows that the scatter around the
line $\rho_{j,l - igm}= \rho_{j,l - dm}$ is little smaller for
smaller $j$. That is, the discrepancy between the IGM and dark matter is
smaller on larger scales. Nevertheless, the discrepancy is still
substantial on scale $j=5$ or 1.03 h$^{-1}$ Mpc, which is larger
than the Jeans length of the IGM.

\begin{figure}[t]
\epsscale{.85}
\plotone{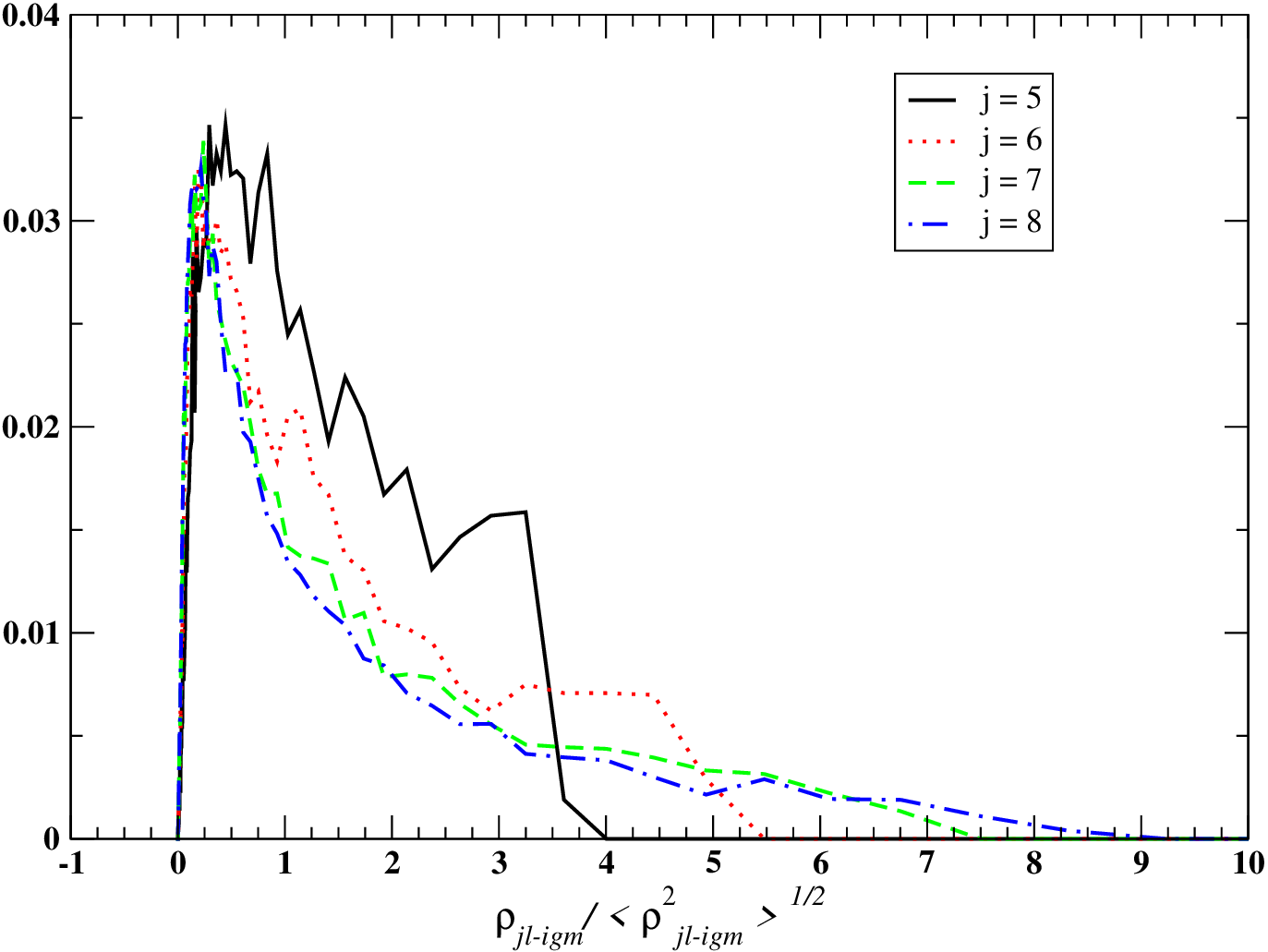}
\caption{One-point distributions of the IGM density
$\rho_{jl-igm}/\langle\rho^2_{jl-igm}\rangle^{1/2}$ at redshift $z=1$ and
on scales $j= 5$, 6, 7, 8. The comoving
scale corresponding to $j$ is $33/2^{j}$ h$^{-1}$ Mpc.}
\end{figure}

When the ``fair sample hypothesis'' (Peebles 1980) holds,
each set of  2$^j$ ($l=0...2^{j-1}$) points  form an ensemble from which the
one-point distribution can be studied.  Figure 3 gives the one-point
distribution of $\rho_{j,l - igm}$
on scale $j=$ 5, 6, 7 and 8 at $z=1$. These distributions are generally
non-Gaussian, having a longer tail at higher $j$, i.e., smaller scales.
These distributions show significant $j$-dependence.
The distribution at $j=8$ has a tail longer than 9 times the
variance, while the tail of the distribution of $j=5$ is only about
four times of the variance. We also see from Fig. 3 that the distribution
has large change from $j=5$ to $j=6$, and from $j=6$ to $j=7$,
but a smaller change from $j=7$ to $j=8$. This is because the scale $j=8$ is
0.129 h$^{-1}$ Mpc which is close to the Jeans length. On these scales
the perturbations in the IGM field are weak.  We show the
redshift-evolution of the $\rho_{j,l - igm}$ one-point distributions
in Fig. 4. The one-point distributions also undergoes  significant
redshift evolution. However, a long tail PDF is already
pronounced at redshift $z=4$.
\begin{figure}[h]
\plotone{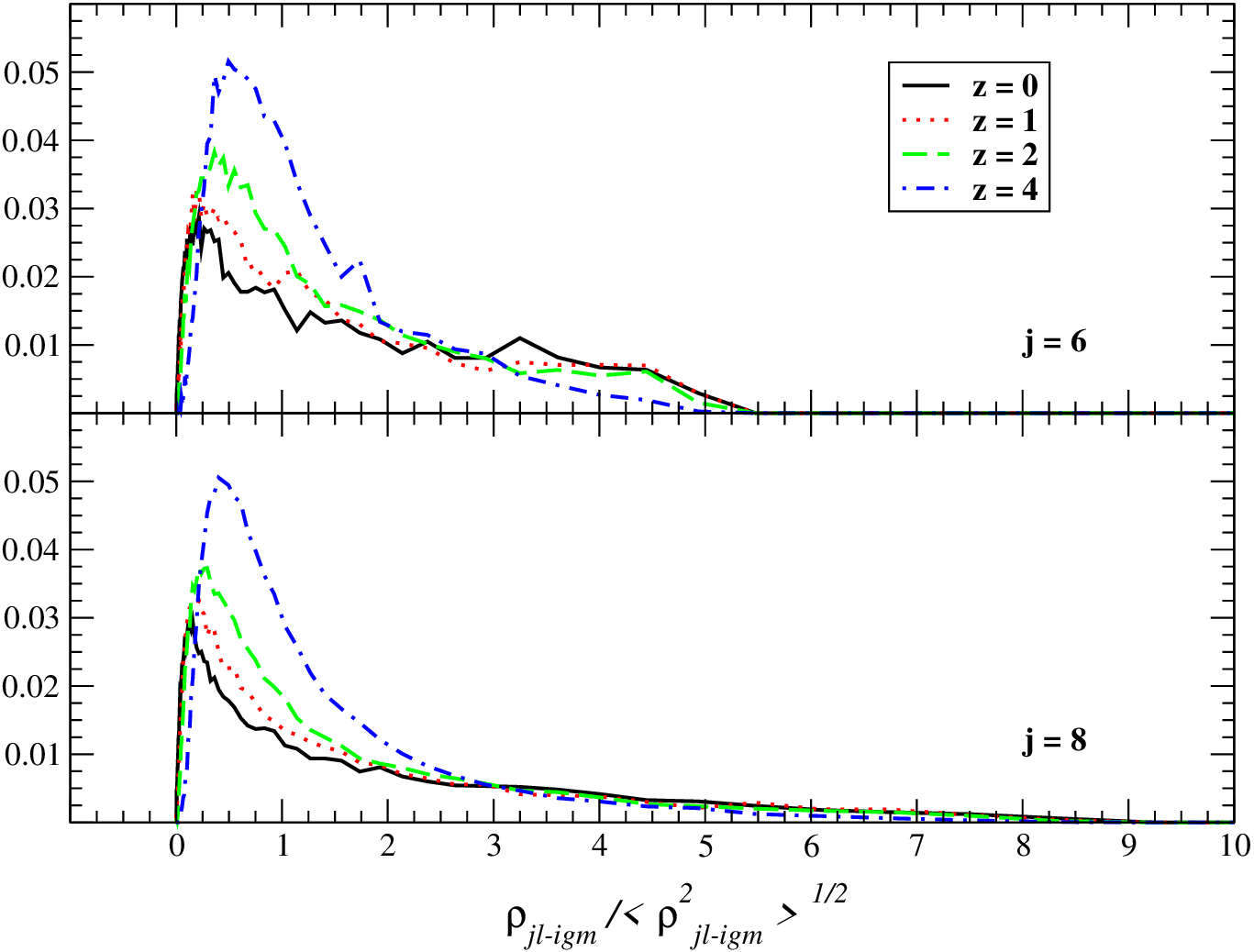}
\caption{The redshift-dependence of the one-point distribution of
the IGM density $\rho_{jl-igm}/\langle\rho^2_{jl-igm}\rangle^{1/2}$
on scales $j=6$ and 8 corresponding to comoving length scales 0.52 and
0.13 h$^{-1}$ Mpc }
\end{figure}

\section{Statistical discrepancy between the IGM and dark matter}

\subsection{One-point distributions of velocity fields}

To demonstrate the statistical discrepancy addressed in \S 2, we first
analyze the one-point distributions of the 1-D velocity fields
$v_{igm}({\bf x})$ and $v_{dm}({\bf x})$. For the linear
solution $v_{igm}({\bf x})=v_{dm}({\bf x})$ [eq.(1)], we have
$v_{j,l -igm}=v_{j,l -dm}$, which are the one-point variables given by
eq.(20) replacing $F(x)$ by the velocity distribution.
It is clear from Figure 5 that the one-point distributions
of $v_{j,l -igm}$ and $v_{j,l -dm}$ at redshift $z=0$ are very different
on all scales, $j=5$, 6, 7 and 8. This difference is
smaller at higher redshift. At $z=4$, the one-point distributions of
both $v_{j,l -igm}$ and $v_{j,l -dm}$ are basically the same on all
scales.

\begin{figure}[h]
\plottwo{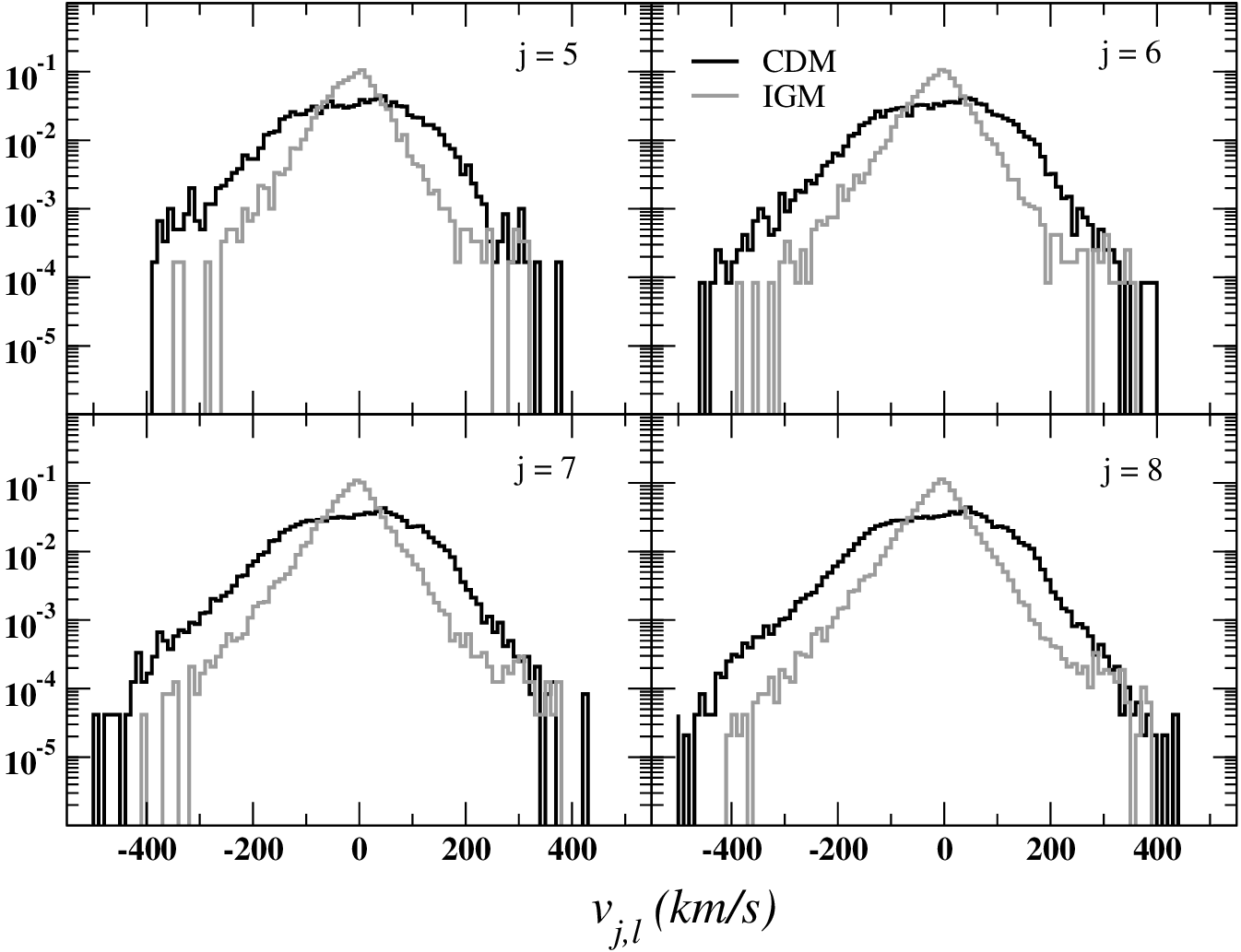}{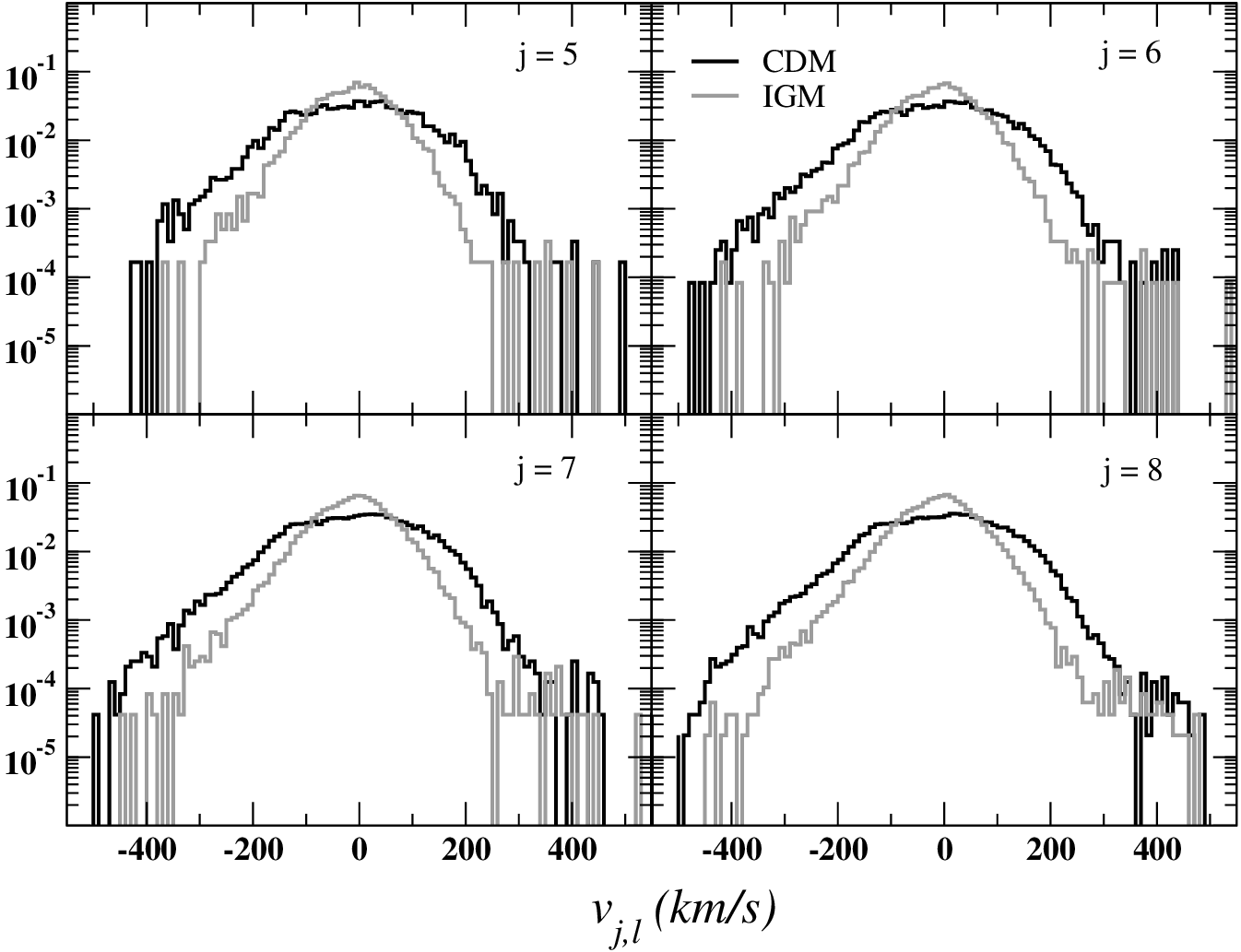}
\end{figure}
\begin{figure}[h]
\plottwo{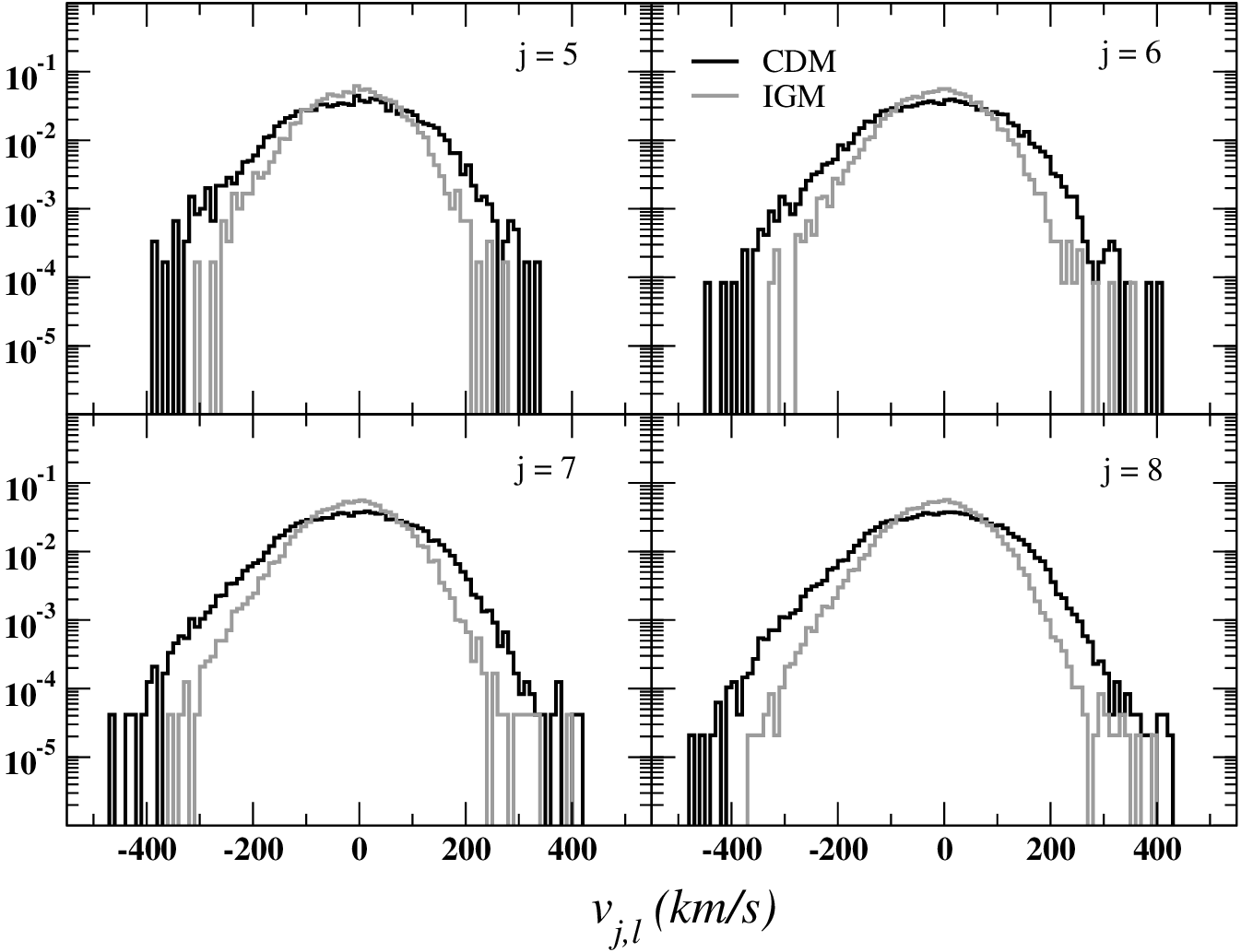}{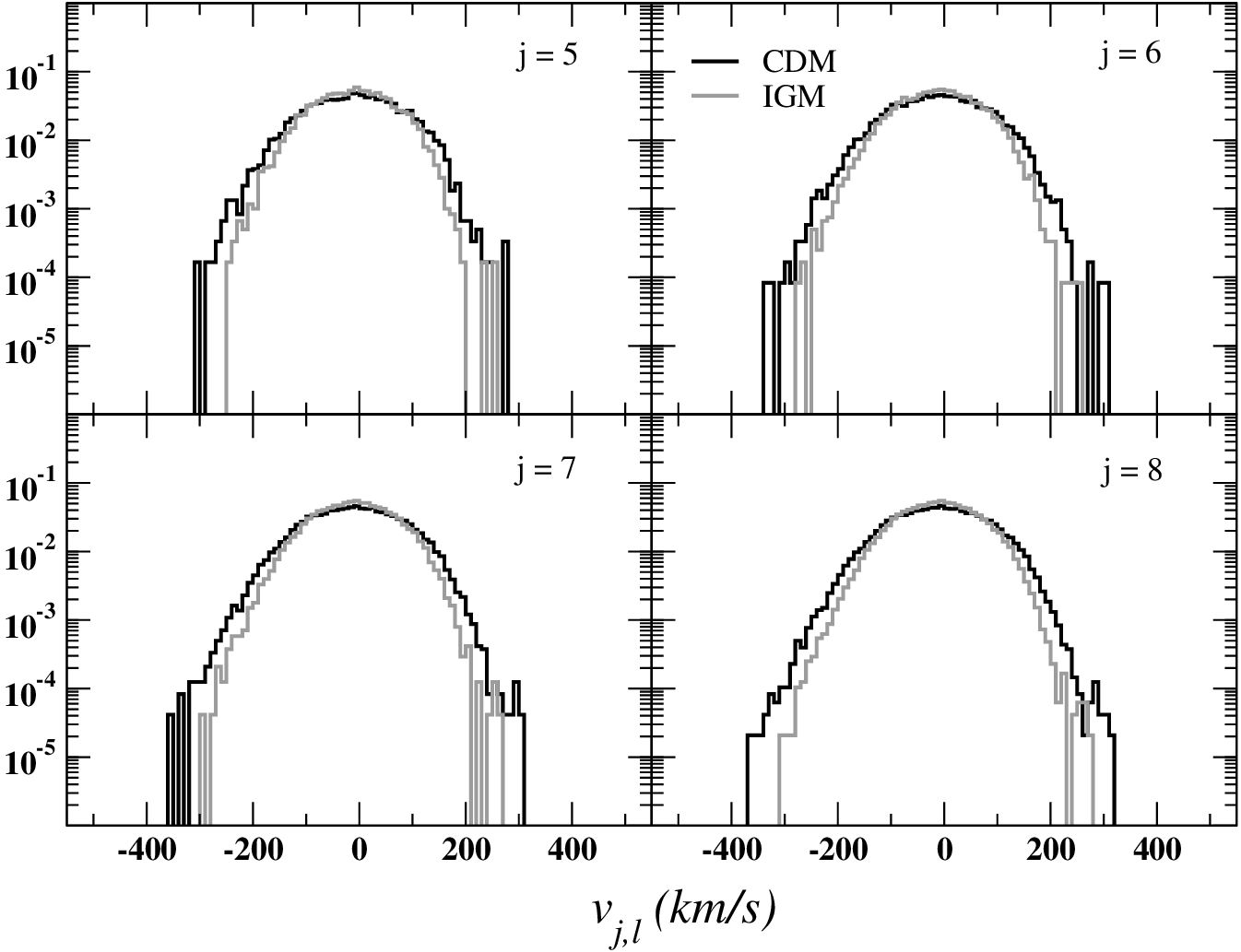}
\caption{One-point distributions of the velocity fields of the IGM
$v_{jl-igm}$ and dark matter $v_{jl-igm}$ on scales $j=5$, 6, 7, 8,
and for redshifts $z=0$ (5a), 1 (5b), 2 (5c), 4 (5d).}
\end{figure}
The nature of the statistical difference between the
$v_{j,l -igm}$ and $v_{j,l -dm}$ one-point distributions
is not only quantitative, but also qualitative.  The $v_{dm}({\bf x})$
one-point distributions are small deviations from a Gaussian PDF
on all scales and all redshifts considered. This result is consistent with
previous studies using N-body simulation samples
(e.g. Yang et al 2001). On the other hand, the one-point distributions of
the IGM velocity field shown in Fig. 5 are generally exponential at
redshifts $z<2$. Even when the scale is as large as
$j=5$ or $\sim 1$ h$^{-1}$ Mpc, the distribution of $v_{igm}({\bf x})$
is still exponential (Fig. 5a). This emphatically shows that the statistical
discrepancy between the IGM and dark matter developed with dynamical
evolution.

\begin{figure}[h]
\plottwo{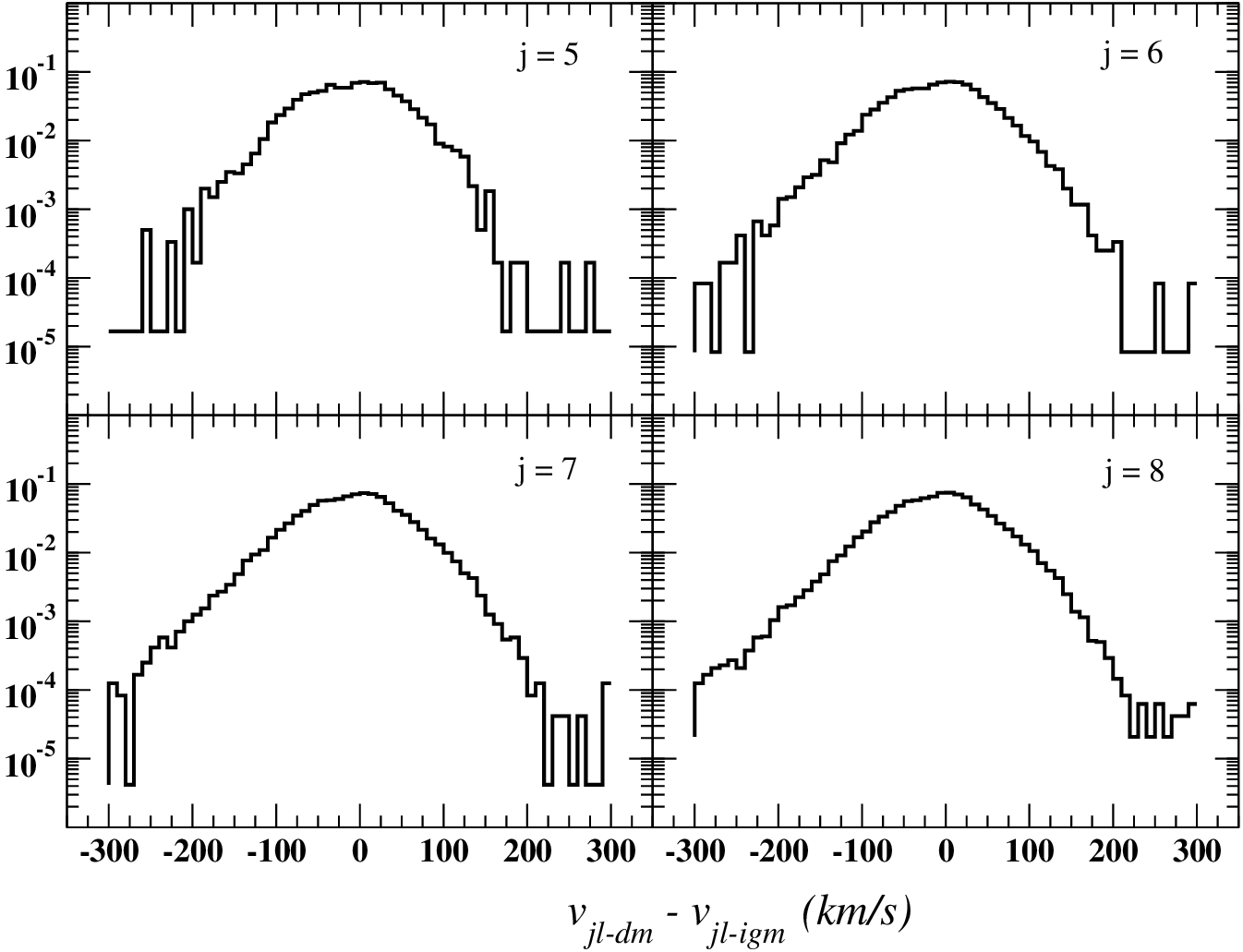}{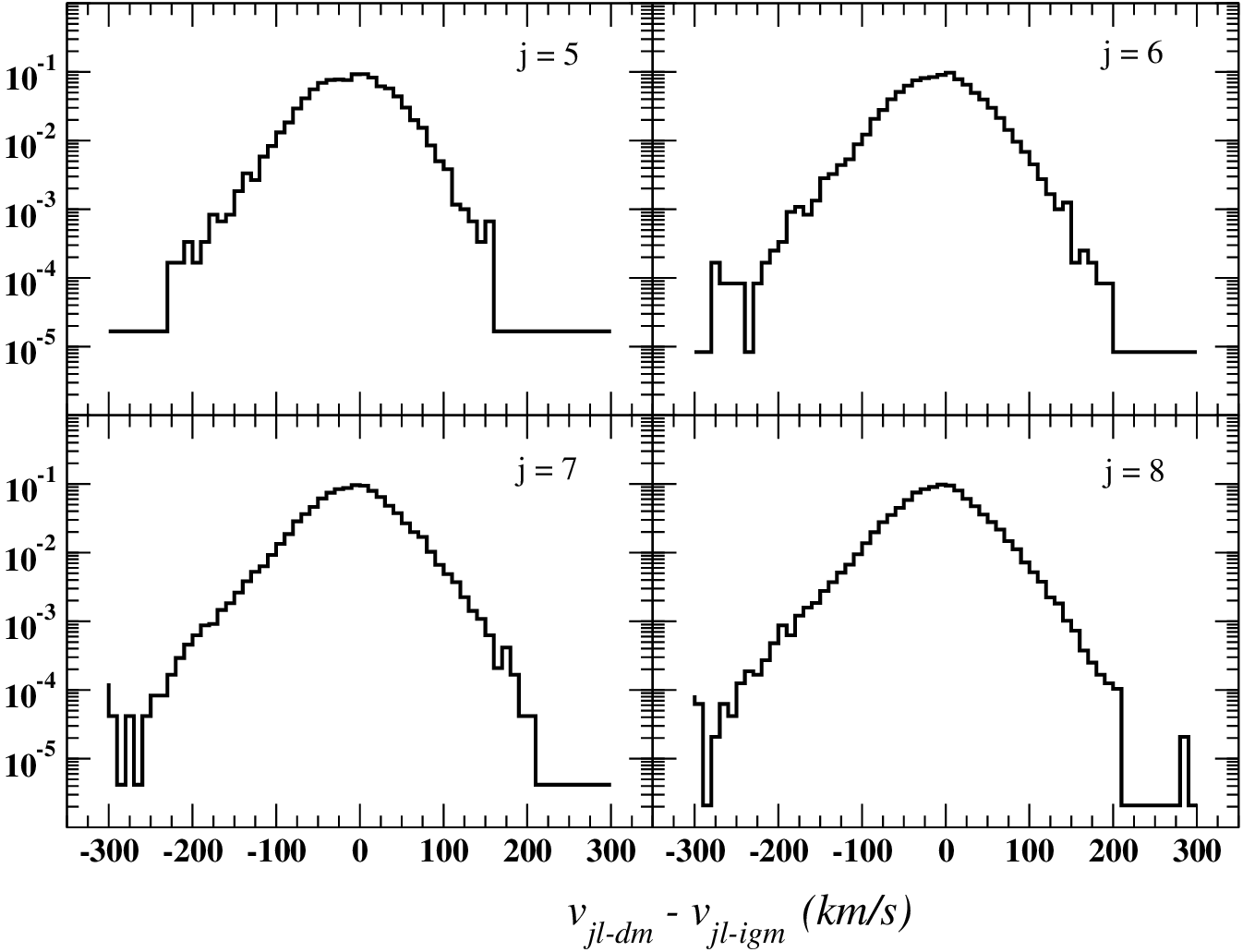}
\end{figure}
\begin{figure}[h]
\plottwo{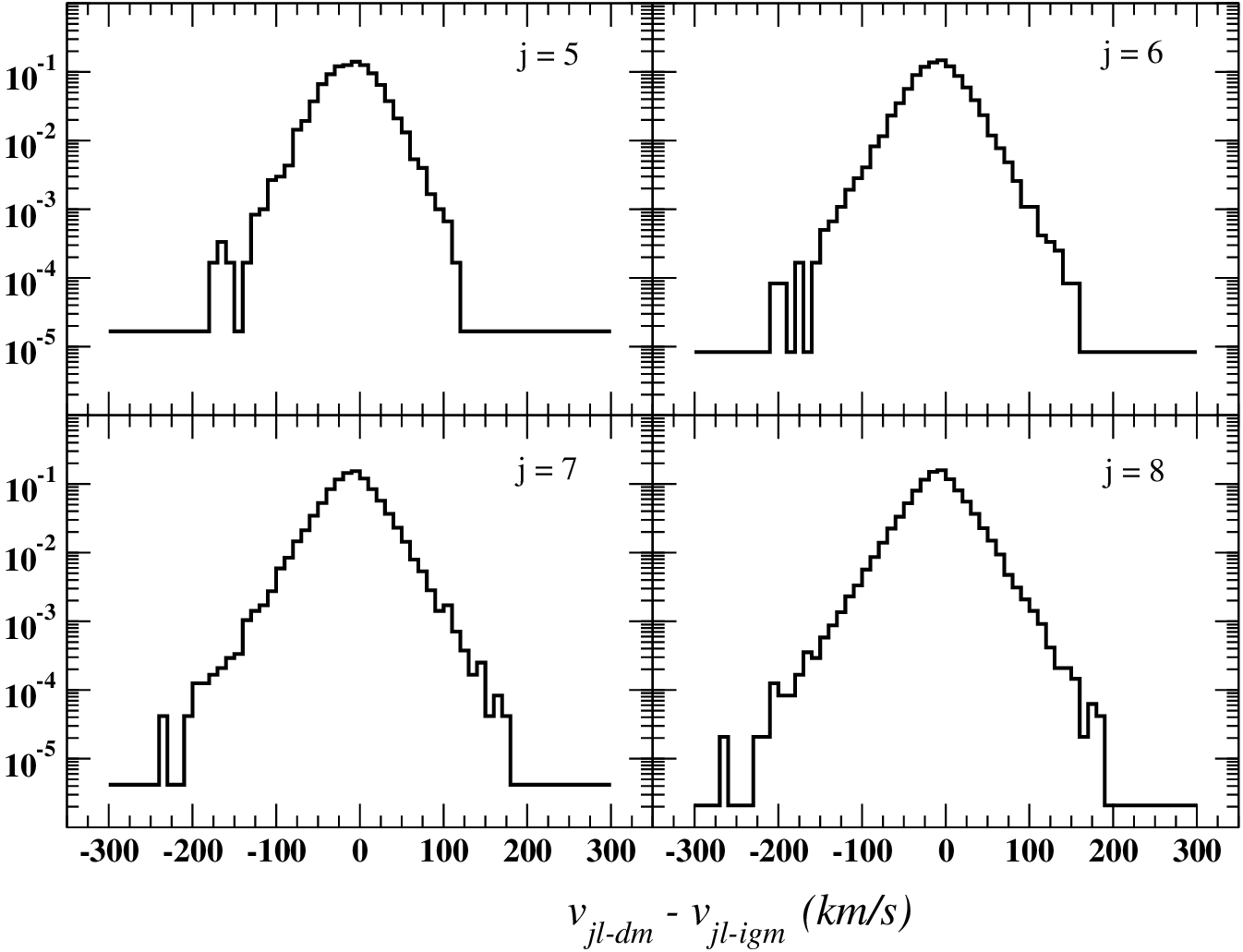}{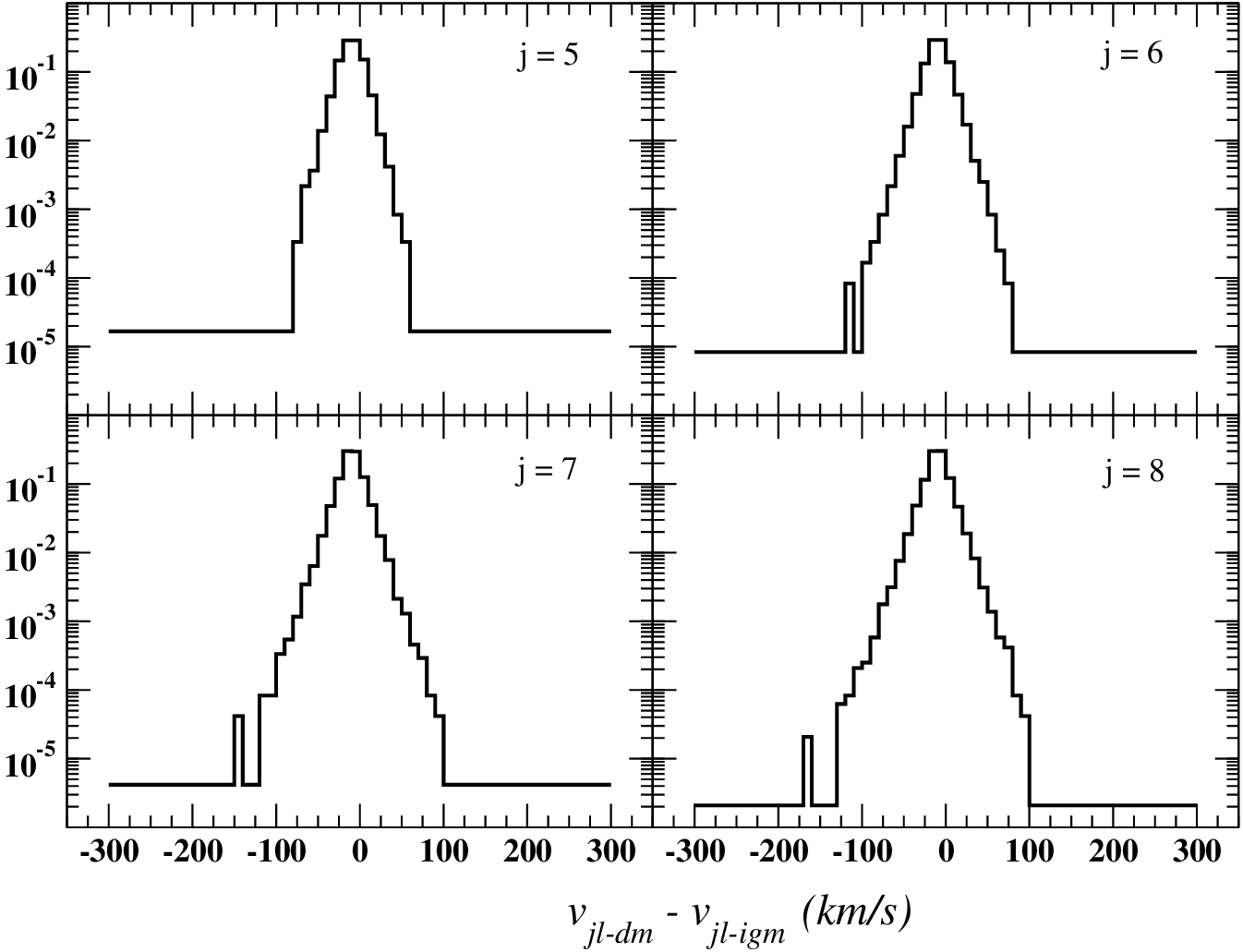}
\caption{One-point distributions of the difference
$v_{jl-dm}-v_{jl-igm}$ on scales $j=5$, 6, 7, 8,
and for redshifts $z=0$ (6a), 1 (6b), 2 (6c), 4 (6d).}
\end{figure}

The dynamical equation (3) for the dark matter looks very similar to
eq.(4) for the IGM.  So why do the two one-point distributions have such
different shapes? The reason is that the $\phi$ term of eq.(4) is truly
an external force, as the gravity potential is independent of the
IGM density and velocity, dependent only on the dark matter.
Therefore, when its Reynolds number is large at lower redshifts, shocks
or Burgers' turbulence will develop in the IGM field due to the external
driving force and the IGM velocity field will be highly non-Gaussian. On
the other hand, the dark matter mass density in eq.(3) is not independent
of the gravitational potential and there is no external driving force. 
Thus, when the gravity potential 
$\phi$ is Gaussian in weakly non-linear evolution, the dark matter 
velocity PDF will approximately be Gaussian too.

The statistical discrepancy of the velocity fields can also be seen with
Fig. 6 which gives the one-point distribution of the difference
$\bigtriangleup{\bf v}_{j,l}\equiv v_{j,l -dm}-v_{j,l -igm}$ on scales
$j=5$, 6, 7 and 8, and redshifts 0, 1,2 and 4. These
one-point distributions are non-Gaussian on all scales and redshift
considered. Although the one-point distributions of $v_{dm}({\bf x})$
and $v_{igm}({\bf x})$ at redshift $z=4$ are not very different
[Fig. 5d], their differences [Fig. 6d] are highly
non-Gaussian. Therefore, the discrepancy between $v_{j,l -dm}$ and
$v_{j,l -igm}$ is not due to noise, but arises from the non-linear evolution
of the IGM fluid.

\subsection{One-point distributions of density fields}

\begin{figure}[h]
\plotone{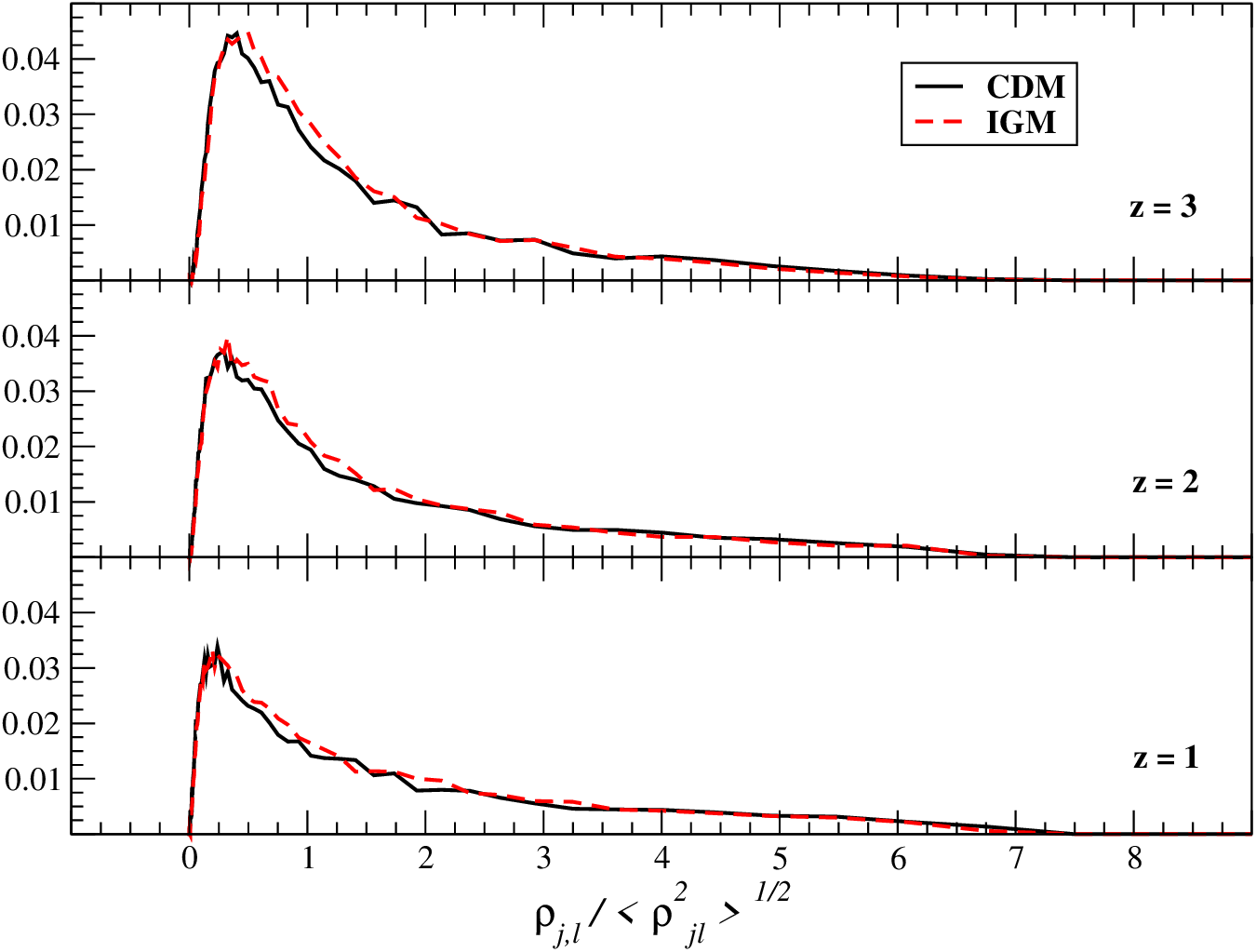}
\caption{One-point distributions of the IGM density
$\rho_{jl-igm}/\langle\rho^2_{jl-igm}\rangle^{1/2}$ and dark matter
$\rho_{jl-dm}/\langle\rho^2_{jl-dm}\rangle^{1/2}$ at redshift
$z=1$, 2 and 3 and on scales $j= 7$, or  comoving length scale
0.26 h$^{-1}$ Mpc.}
\end{figure}

Figure 7 presents the one-point distributions for
$\rho_{j,l - igm}$ and $\rho_{j,l - dm}$ on scale $j=7$ and redshifts
$z=1$, 2 and 3. The horizontal axes are the variance-normalized densities,
$\rho_{j,l - igm}/\langle \rho^2_{j,l - igm}\rangle^{1/2}$ and
$\rho_{j,l - igm}/\langle \rho^2_{j,l - dm}\rangle^{1/2}$. Unlike the 
velocity fields, both the IGM and dark matter show highly non-Gaussian 
behavior even at redshift $z=3$. Additionally, the two curves
in Fig. 7 are not significantly different. That is, the mass density 
one-point distribution is not as sensitive to the statistical discrepancy 
as the velocity one-point distribution. This is probably because the 
statistical discrepancy
between a ``passive substance" and underlying field is significant only for
non-conserved quantities (temperature, velocity etc.) (Shraiman \& Siggia
2001.)  Nevertheless, one can
see from Fig. 7 that the IGM distributions are always a little higher than
the corresponding distribution for the dark matter at the range around
$\rho_{j,l - igm}/\langle \rho^2_{j,l - igm}\rangle^{1/2}\simeq 1$. This
indicates that the tail of the IGM one-point distribution should be
shorter than that of dark matter. Therefore, we can expect that the high
order moments of the one-point distributions will more clearly show the
discrepancy between the IGM and dark matter.

In this paper, we use only the second order moment. The correlation between
$\rho_{j,l - igm}$ and $\rho_{j,l - dm}$ can be measured with the ratio
 $I_{dm}$, $I_{igm}$ [eqs.(10), (11)], or $I$ [eq.(12)]. Using the DWT
one-point variables $\rho_{j,l - igm}$ and $\rho_{j,l - dm}$, we can
redefine, respectively, the ratios $I_{dm}$ and $I_{igm}$ as
\begin{equation}
I_{j, -dm}=\frac{[\langle[\rho_{j,l - dm}-\rho_{j,l - igm}]^2\rangle^{1/2}}
    {\langle \rho_{j,l - dm}^2\rangle^{1/2}}
\end{equation}
\begin{equation}
I_{j, -igm}=   \frac{[\langle[\rho_{j,l - dm}-
     \rho_{j,l - igm}]^2\rangle^{1/2}}
    {\langle \rho_{j,l - igm}^2\rangle^{1/2}}
\end{equation}

\begin{figure}[t]
\plotone{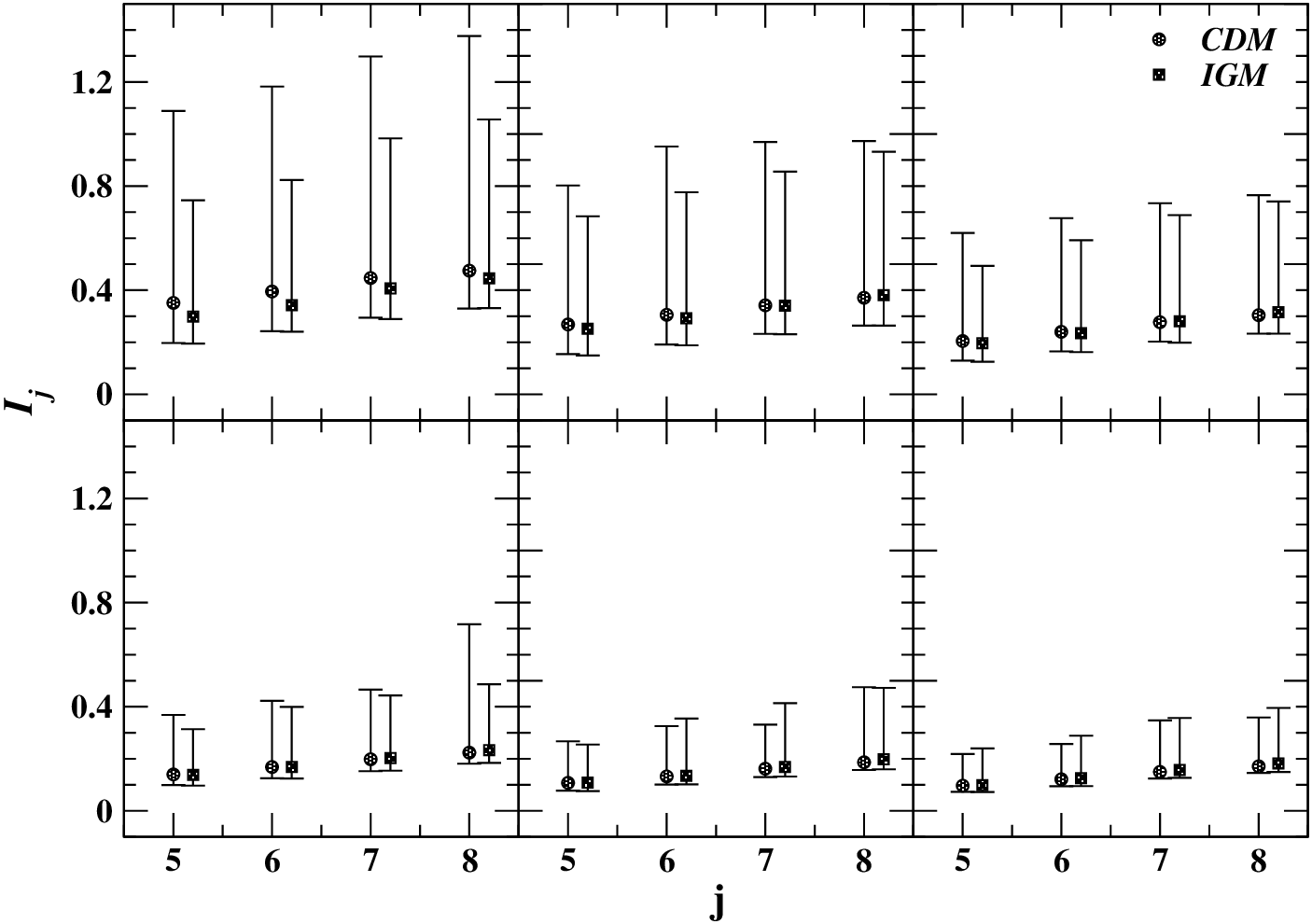}
\caption{The ratio $I_j$ at redshifts $z=0$, top left panel, to
redhsifts 0.5, 1, 2, 3, 4, (moving left to right) and
on scales $j=5$, 6, 7, 8, which corresponds to the comoving scales
$33/2^{j}$ h$^{-1}$ Mpc.}
\end{figure}

 From the one-point distributions of Fig. 7, we have
$\langle \rho_{j,l - dm}^2\rangle \simeq \langle \rho_{j,l - igm}^2\rangle$.
Thus, if $\rho_{j,l - igm}$ perfectly traces $\rho_{j,l - dm}$, we have
correlation $\langle\rho_{j,l - igm}\rho_{j,l - dm}\rangle \simeq
\langle\rho^2_{j,l - igm}\rangle \simeq \langle \rho^2_{j,l - dm}\rangle$,
and therefore, $I_{j, -dm} \simeq I_{j, -igm} \simeq 0$. If
$\rho_{j,l - igm}$ is fully independent of $\rho_{j,l - dm}$, we have
$I_{j, -dm}\simeq I_{j, -igm} \simeq \sqrt{2}$.

Figure 8 plots $I_{j, -dm}$ and $I_{j, -igm}$ for $z=0$, 0,5, 1, 2, 3 and 4
on scales $j=5$, 6, 7 and 8. The error bars are the 67\% confidence level
of the 500 1-D samples. Figure 8 shows that $I_j$ is always larger
than 0, equal to $\sim 0.1 - 0.2$ at redshift $\geq 2$, and
$\sim 0.3 - 0.6$ at redshift $<2$, and weakly dependent on scales. That is,
the cross correlation at redshift $<$ 2 is
\begin{equation}
\frac{\langle\rho_{j,l - igm}\rho_{j,l - dm}\rangle}
{\langle\rho^2_{j,l - igm}\rangle}, {\rm \ \ \ or \ \ \ }
\frac{\langle\rho_{j,l - igm}\rho_{j,l - dm}\rangle}
{\langle\rho^2_{j,l - dm}\rangle} \simeq 0.70 - 85
\end{equation}
Therefore, the IGM is not a perfect tracer at low redshift and on scales
larger than the Jeans length. These results indicate that
15 -30\% of the two density fields are off-correlated. This is consistent
with the analysis in \S 2.2.

\subsection{One-point distribution of $\bigtriangleup \delta({\bf x})$}

\begin{figure}[t]
\plotone{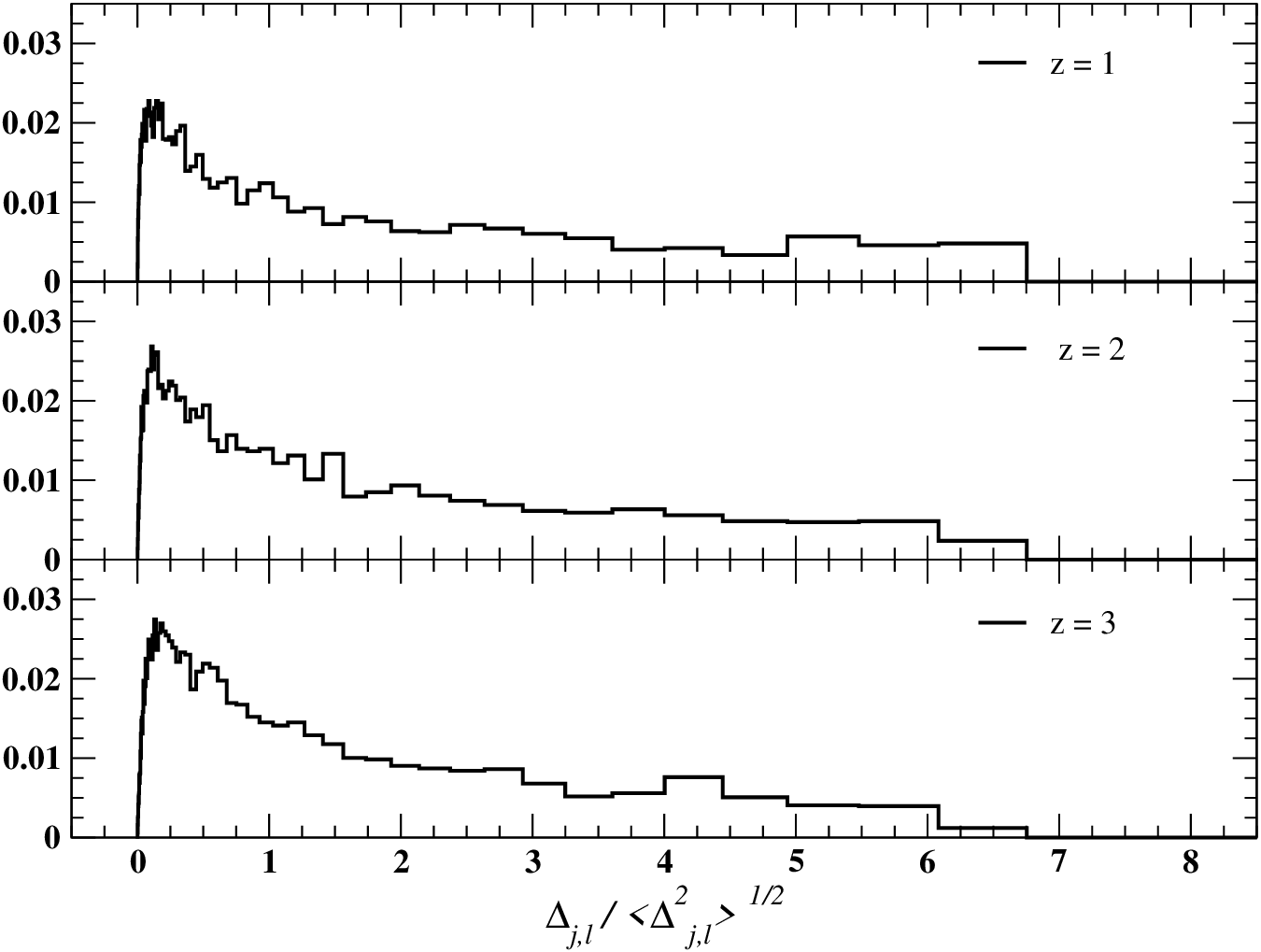}
\caption{One-point distributions of the difference
$\bigtriangleup_{j,l}/\langle\bigtriangleup^2_{j,l}\rangle^{1/2}> 0$
 for redshift $z=1$, 2 and 3,
and on scale $j=7$, or 0.26 h$^{-1}$ Mpc.}
\end{figure}

The DWT one-point variable for the density difference
$\bigtriangleup \delta(x)$ [eq.(2)] is given by
\begin{equation}
\bigtriangleup_{j,l}=\int
  \bigtriangleup \delta(x)\phi_{j,l}(x)dx.
\end{equation}
Since $\phi_{j,l}(x)$ is localized in cell $(j,l)$, the integral
$\int \phi_{j,l}\phi_{j,l'}\phi_{j,l''}dx$ is generally  non-zero
only for $l=l'=l''$ (for Haar wavelet, this is exactly true. For other
wavelets, the integral with $l\neq l',\ l''$ is much less than
that of $l=l'=l''$). Thus, for two fields $A(x)$ and $B(x)$,
we have $\int A(x)B(x)\psi_{jl}(x)dx \simeq g_j \epsilon^A_{jl}
\epsilon^B_{jl}$, where $\epsilon^A_{jl}$ and $\epsilon^B_{jl}$
are the WFCs of $A(x)$ and $B(x)$. The factor $g_j$ is
\begin{equation}
g_j=\int \psi^3_{jl}(x)dx.
\end{equation}
which is the factor $g_W$ used in eq.(16).

With $\bigtriangleup_{j,l}$, eq. (15) can be rewritten as
\begin{equation}
\frac{d\bigtriangleup_{j,l}}{d t}= -
\frac{1}{a} \lambda_{j}\bigtriangleup_{j,l} + \frac{1}{a}\eta_{j},
\end{equation}
where
\begin{equation}
\eta_j = \int dx'\psi_{jl}(x')
(\nabla \cdot \bigtriangleup {\bf v}(x)).
\end{equation}
and
\begin{equation}
\lambda_j = g_j\int x'\psi_{jl}(x')
  (\nabla \cdot {\bf v}_{dm}).
\end{equation}
As discussed in \S 2.3, the PDF of $\bigtriangleup_{j,l}$ should be long 
tailed.

The normalized one-point distributions of
$\bigtriangleup_{j,l}/\langle\bigtriangleup^2_{j,l}\rangle^{1/2}$
for $j=7$ and redshifts 1,2, and 3 are plotted in Fig. 9. All the tails of
the distributions are remarkably long. They have 6 $\sigma$ events
at $z \leq 3$ with probability $\sim$ 0.005, while for a Gaussian field,
the probability of 6-$\sigma$ is $ < 10^{-8}$. This shows again that
the statistical discrepancy between the IGM and dark matter is not due
to  processes like Jeans diffusion or Gaussian noise. Fitting the tails 
with a power law
$(\bigtriangleup_{j,l}/\langle\bigtriangleup^2_{j,l}\rangle^{1/2})^{-\beta}$,
we have $\beta < 1$.  This result is consistent with the dynamical 
eq.(28).

\section{Conclusion}

Using the dynamical equations governing the IGM, we have shown that the
decoupling of important statistical properties of the gas component from
the underlying dark matter field on scales larger than the Jeans length
is inevitable. The one-point distribution of the IGM peculiar velocity
field is found to be substantially different from that of dark matter
at redshift $\leq 2$. Although the one-point distribution of the IGM
mass density field shows only a little deviation from that of the dark 
matter, the density fields of the IGM and dark matter are not
point-by-point correlated, and a significant part of the two fields is
off-correlated on all redshifts $z\leq 4$.

It is difficult to explain the discrepancy as an effect of Jeans 
smoothing, as the discrepancy is still evident even when the fields
are smoothed on scales $>1$ h$^{-1}$ Mpc ($j=5$).  This is much larger
than the Jeans length of the IGM with temperature $\sim 10^4$ K at
the mean density. The Jeans length of the IGM may be $>1$ h$^{-1}$ Mpc
in an area with high temperatures $\simeq 10^7$ K, but not one with high 
density. Yet, even in this case Jeans smoothing does not adequately explain 
the statistical discrepancy, because the difference between the density 
distributions of the IGM and dark matter is highly non-Gaussian.

The decoupling of the IGM from the dark
matter mass field is important for problems in large scale structure 
formation. Some observations have already implied that the IGM and dark
matter have decoupled. For instance, X-ray measurements find no
evidence for the baryon fraction of clusters to be equal to the universal 
value from cosmological nucleosynthesis (White \& Fabian 1995; David
1997; Ettori, Fabian \& White 1997; White, Jones \& Forman 1997; 
Ettori \& Fabian, 1999). This result violates the similarity 
the IGM and dark matter if typical galaxy clusters formed from linear 
fields on scales of a few ten (comoving) Mpc, which is much larger than 
the Jeans length of the IGM. Moreover, the X-ray observed 
luminosity-temperature relation for groups and clusters is found to be 
inconsistent with the prediction given by the similarity between the 
IGM and dark matter. The high entropy floor observed in nearby groups 
and low mass clusters directly violates the dynamical self-similar scaling 
(Ponman, Cannon, \& Navarro 1999; Lloyd-Davies, Ponman, \& Cannon 2000). 
Although, at present we cannot attribute all these discrepancies to the 
decoupling shown in this paper, we now understand that the dynamics of 
the IGM and dark matter fields will lead to a qualitatively different 
evolution of those fields. While we did not consider stellar formation 
and its feedback to the IGM in our present simulations, we believe that 
the statistical discrepancy will still be a common feature even when 
these effects are considered because the reason for discrepancy we have 
uncovered arises from the nonlinear evolution of the random fields of 
the IGM and dark matter.

Finally, we should mention the effect of the size of the simulated sample.
Since the simulation box is only 25 h$^{-1}$ Mpc, the
use of the fair sample hypothesis may not be appropriate when considering
perturbations on scales larger than this. We can estimate the           
effect of these larger perturbations by adding long-wavelength                
modes (Tormen \& Bertscinger 1996). Since the IGM and dark matter fields
are linear or quasi-linear on scales larger than 25 h$^{-1}$  
Mpc, the effect of long-wavelength perturbations can be analyzed by
adding a displacement to each particle in the simulation box.
The displacement is given by the linear field or Zeldovich approximation
of the field consisting of modes on scales larger than 25 h$^{-1}$ Mpc. As
was discussed in \S 1, there is no discrepancy between the IGM and
dark matter in the linear or quasi-linear regime. Thus, the displacement for
the IGM is the same as that of the dark matter meaning that long wavelength 
modes are not a source of the discrepancy between the IGM and dark matter.

\acknowledgments

We thank Drs. C.W Shu and P. He for their help during the preparation
of this revised version. Thanks also to our referee for his(her) comments. LLF 
acknowledges support from the National Science Foundation of China (NSFC) 
and National Key Basic Research Science Foundation.

\appendix

\section{Hydrodynamic equations for dark matter fields}

Let us consider a flat universe having cosmic scale factor $a(t)\propto
t^{2/3}$ and dominated by dark matter. In a hydrodynamic description,
the dark matter is described by a mass density field $\rho_{dm}({\bf x}, t)$
and a peculiar velocity field ${\bf v}_{dm}({\bf x}, t)$, where
${\bf x}$ is the comoving  coordinate. The field is described by the
equations of continuity, momentum and gravitational potential as
(Wasserman 1978)
\begin{equation}
\frac{\partial \delta_{dm}}{\partial t} +
  \frac{1}{a}\nabla \cdot (1+\delta_{dm}) {\bf v}_{dm}=0
\end{equation}
\begin{equation}
\frac{\partial a{\bf v}_{dm}}{\partial t}+
 ({\bf v}_{dm}\cdot \nabla){\bf v}_{dm}= -\nabla \phi
\end{equation}
\begin{equation}
\nabla^2 \phi = 4\pi G a^2\bar{\rho}_{dm}\delta_{dm}.
\end{equation}
  The mean
density is $\bar{\rho}_{dm}(t) =1/6\pi Gt^2 \propto a^{-3}$. The
gravitational potential $\phi$ is zero (or constant) when the density
perturbation $\delta_{dm}=0$. The operator $\nabla$ acts on the comoving
coordinate ${\bf x}$.

For growth modes in the perturbations, velocity is irrotational. We can
then define a velocity potential by
\begin{equation}
{\bf v}_{dm}=- \frac {1}{a}\nabla \varphi_{dm}.
\end{equation}
The momentum equation (A2) can then be rewritten as
\begin{equation}
\frac{\partial \varphi_{dm}}{\partial t}-
\frac{1}{2a^2}(\nabla \varphi_{dm})^2 = \phi.
\end{equation}
This is the Bernoulli equation.

\section{Hydrodynamic equations for the IGM}

As usual, the IGM is assumed to be ideal fluid with polytropic index
$\gamma=5/3$. The hydrodynamic equations of the IGM are (Peebles 1980)
\begin{equation}
\frac{\partial \delta_{igm}}{\partial t} +
  \frac{1}{a}\nabla \cdot (1+\delta_{igm}) {\bf v}_{igm}=0
\end{equation}
\begin{equation}
\frac{\partial a{\bf v}_{igm}}{\partial t}+
 ({\bf v}_{igm}\cdot \nabla){\bf v}_{igm}=
-\frac{1}{\rho_{igm}}\nabla p - \nabla \phi
\end{equation}
\begin{equation}
\frac{\partial {\cal E}}{\partial t}+5\frac{\dot{a}}{a}{\cal E}+
  \frac{1}{a}\nabla\cdot ({\cal E}{\bf v}_{igm})=
   -\frac{1}{a}\nabla\cdot(p{\bf v}_{igm})-
   \frac{1}{a}\rho_{igm}{\bf v}_{igm}\cdot \nabla\phi- \Lambda_{rad},
\end{equation}
where $\rho_{igm}$, ${\bf v}_{igm}$, ${\cal E}$ and $p$ are, respectively,
the mass density, peculiar velocity, energy density and pressure of the IGM.
The term $\Lambda_{rad}$ in Eq.(2) is given by the radiative
heating-cooling of the baryonic gas per unit volume. The gravitational
potential $\phi$ in eqs.(B2) and (B3) can still be given by eq. (A3).
That is, the gravity of the IGM is negligible. The evolution of the IGM
mass field $\rho_{igm}({\bf x},t)$ is governed by the gravity of dark
matter only.

The hydrodynamic equations for the IGM, eqs.(B1)-(B3) can be written
in the form of conservation laws for mass, momentum, and energy in a
comoving volume as
\begin{equation}
\frac {\partial a^3\rho}{\partial t}+ \frac{1}{a}
   \frac{\partial}{\partial x_i}(a^3\rho v_i)=0,
\end{equation}
\begin{equation}
\frac{\partial a^3\rho v_i}{\partial t}+
\frac{1}{a}\frac{\partial}{\partial x_j}(a^3\rho v_iv_j +a^3p\delta_{ij})
  = -\dot{a}a^2\rho v_i - a^2\rho \nabla \phi
\end{equation}
\begin{equation}
\frac{\partial a^3{\cal E}}{\partial t}+
\frac{1}{a}\frac{\partial}{\partial x_i}[a^3({\cal E}+p)v_i]=
   -2\dot{a}a^2{\cal E}
  -a^2\rho {\bf v}\cdot \nabla \phi - a^3\Lambda_{rad}.
\end{equation}
In eqs.(B4)-(B6), we dropped the subscript $igm$ for simplicity.

To sketch the gravitational clustering of the IGM, it is not necessary to
consider the details of heating and cooling. Thermal processes are generally
highly localized, and therefore, it is reasonable to describe all thermal
processes by a polytropic relation
$p({\bf x},t) \propto \rho_{igm}^{\gamma}({\bf x},t)$. Thus eq.(B2) becomes
\begin{equation}
\frac{\partial a{\bf v}_{igm}}{\partial t}+
 ({\bf v}_{igm}\cdot \nabla){\bf v}_{igm}=
  -\frac{\gamma k_B T}{\mu m_p} \frac{\nabla \delta_{igm}}{(1+\delta_{igm})}
  - \nabla \phi
\end{equation}
where the parameter $\mu$ is the mean molecular weight of the IGM
particles, and $m_p$ the proton mass. Here, we don't need the
energy equation and the IGM temperature evolves as
$T \propto \rho^{\gamma-1}$, or $T =T_0(1+\delta_{igm})^{\gamma-1}$.

Eq.(B7) differs from eq.(A2) only by the temperature-dependent term. If
we treat this term in the linear approximation, we have
\begin{equation}
\frac{\partial \varphi_{igm}}{\partial t}-
\frac{1}{2a^2}(\nabla \varphi_{igm})^2 -
\frac{\nu}{a^2}\nabla^2 \varphi_{igm}
=\phi,
\end{equation}
where $\varphi_{igm}$ is the velocity potential for the IGM field defined by
\begin{equation}
{\bf v}_{igm}= - \frac {1}{a}\nabla \varphi_{igm}.
\end{equation}
The coefficient $\nu$ is given by
\begin{equation}
\nu=\frac{\gamma k_BT_0}{\mu m_p (d \ln D(t)/dt)},
\end{equation}
where $D(t)$ describes the linear growth behavior. The term with $\nu$ in
eq.(B8) acts like a viscosity (due to thermal diffusion)
characterized by the Jeans length $k_J^2=(a^2/t^2)(\nu m_p/\gamma k_BT_0)$.

The unperturbed solutions of the density and velocity fields of both
dark and baryonic matter are
$\bar{\rho}_b=(\Omega_b/\Omega_{dm})\bar{\rho} \propto a^{-3}$, and
${\bf v}_b= {\bf v}=0$, where $\Omega_b$ and $\Omega_{dm}$ are, respectively, 
the density parameters of the IGM and dark matter. Therefore, the 
linearlization of eqs. (B1) and (B7) yields
\begin{equation}
\frac{\partial \delta_{igm}}{\partial t} +
  \frac{1}{a}\nabla \cdot {\bf v}_{igm}=0
\end{equation}
\begin{equation}
\frac{\partial a{\bf v}_{igm}}{\partial t} =
-\frac{\gamma k_B \overline{T}}{\mu m_p}\nabla \delta_{igm} - \nabla \phi
\end{equation}
where the mean temperature
$\overline{T}\propto \overline{\rho_b}^{\gamma-1} \propto a^{-3(\gamma-1)}$.
In Fourier space, we have
\begin{equation}
\frac{\partial^2 \delta_{igm}({\bf k},t)}{\partial t^2} +
2\frac {\dot{a}}{a}\frac{\partial\delta_{igm}({\bf k}, t)}{\partial t}+
\frac{1}{t^2}\frac{k^2}{k_J^2}\delta_{igm}({\bf k},t) =
4\pi G\bar{\rho}\delta_{dm}({\bf k},t)
\end{equation}
\begin{equation}
\frac{\partial \upsilon_{igm}({\bf k},t)}{\partial t}+
\frac{\dot{a}}{a}\upsilon_{igm}({\bf k},t)= -
\frac{1}{t^2a}\frac{1}{k^2_J}\delta_{igm}({\bf k},t)
  +\frac {4\pi G \bar{\rho}a}{k^2}\delta_{dm}({\bf k},t),
\end{equation}
where ${\bf v}_{igm}({\bf k},t)=i{\bf k}\upsilon_{igm}({\bf k},t)$, and the 
Jeans wavenumber
$k^2_J=(a^2/t^2)(\mu m_p/\gamma k_b \overline{T}) \propto a^{3\gamma -4}$.
If $\gamma = 4/3$, $k_J$ is time-independent.

In solving equations (B13) and (B14), we consider only the growth mode
of the perturbation of dark matter, i.e. $\delta({\bf k},t) \propto a$.
In the case of $\gamma=4/3$, the solution of eqs.(B13) and (B14) is
(Bi, B\"orner, Chu 1993)
\begin{equation}
\delta_{igm}({\bf k},t) =\frac{\delta_{bm}({\bf k},t)}{1+3k^2/2k_J^2} +
 c_1t^{-(1+\epsilon)/6}+c_2t^{-(1-\epsilon)/6}
\end{equation}
where $\epsilon=(1-4k^2/9k_J^2)^{1/2}$, and constants $c_1$ and $c_2$
depend on the initial condition $\delta_{igm}({\bf k},0)$ and
$\upsilon_{igm}({\bf k},0)$. Therefore, regardless the initial condition of
IGM, after a long evolution we have
\begin{equation}
\delta_{igm}({\bf k},t) = \delta_{dm}({\bf k},t), \hspace{5mm}
{\bf v}_{igm}({\bf k},t) = {\bf v}_{dm}({\bf k},t), \hspace{1cm}
{\rm if} \ k \ll k_J.
\end{equation}
These solutions mean that the initial conditions of the IGM are 
unimportant. The IGM will, in the end, follow the same trajectory as the
dark matter on scales larger than the Jeans length.

The  solutions of linear equations (B11) and (B12) have also been found 
by using assumptions of the IGM thermal processes other than that used 
in eqs.(B13) and (B14) (e.g. Nusser 2000, Matarrese \& Mohayee 2002). 
A common feature of these solutions is
\begin{equation}
\delta_b({\bf k},t) = (1+ {\rm decaying \ terms})\delta({\bf k},t)
  + {\rm decaying \ terms},
  \hspace{1cm} {\rm if} \ k \ll k_J.
\end{equation}
The initial conditions of the IGM field affects only the decaying terms,
and therefore, the linear solutions eq.(B16) hold in general regardless
specific assumptions of IGM thermal processes.

\section{PDF of $\bigtriangleup_{j,l}$ in eq.(27)}

Using $d\tau=dt/a$, eq. (28) can be rewritten as
\begin{equation}
\frac{d \bigtriangleup_{j,l}}{d \tau}= -\lambda_{j}
\bigtriangleup_{j,l}  + \eta_{j}.
\end{equation}
If the stochastic forces $\lambda_{j}$ and $\eta_{j}$ are Gaussian,
then
\begin{eqnarray}
\langle \lambda_{j} \rangle & = & \overline{\lambda}_{j} \\ \nonumber
\langle [\lambda_{j}(\tau)-\overline{\lambda}_{j}]
     [\lambda_{j}(\tau')-\overline{\lambda}_{j}]\rangle &  = &
   2D_{\lambda, j}\delta (\tau-\tau')\\ \nonumber
\langle \eta_{j} \rangle & = & 0 \\ \nonumber
\langle \eta_{j}(\tau)\eta_{j}(\tau')\rangle & = & =2D_{\eta, j} 
\delta(\tau-\tau')
\end{eqnarray}
The Fokker-Planck equation corresponding to eq.(C1) is
(Venkataramani et al. 1996)
\begin{equation}
\frac{\partial }{\partial \tau}P(\bigtriangleup_{j,l}, \tau)=-
   \frac{\partial }{\partial \bigtriangleup_{j,l}} j
(\bigtriangleup_{j,l}, \tau)
\end{equation}
where $P(\bigtriangleup_{j,l}, \tau)$ is the PDF of
$\bigtriangleup_{j,l}$, and
the flux $j(\bigtriangleup_{j,l}, \tau)$ is given by
\begin{equation}
j(\bigtriangleup_{j,l}, \tau)=
(-\overline{\lambda}_{j} +
D_{\lambda, j})\bigtriangleup_{j,l} P(\bigtriangleup_{j,l}, \tau)
-\frac{\partial}{\partial \bigtriangleup_{j,l}}
[D_{\lambda, j}|\bigtriangleup_{j,l}|^2 + D_{\eta, j}]
P(\bigtriangleup_{j,l}, \tau).
\end{equation}

For stationary solution,
$\partial P(\bigtriangleup_{j,l}, \tau)/\partial \tau=0$,
and therefore,
$\partial j(\bigtriangleup_{j,l}, \tau)/\partial\bigtriangleup_{j,l}=0$.
We have $j(\bigtriangleup_{j,l})=$ const. However, when
$\bigtriangleup_{j,l}$ is very large, we should have
$j(\bigtriangleup_{j,l})=0$, and therefore $j(\delta^R_D)=0$. Thus, we
have equation as
\begin{equation}
(-\overline{\lambda}_{j} +
D_{\lambda, j})\bigtriangleup_{j,l} P(\bigtriangleup_{j,l}, \tau)
-\frac{\partial}{\partial \bigtriangleup_{j,l}}
[D_{\lambda, j}|\bigtriangleup_{j,l}|^2 + D_{\eta, j}]
P(\bigtriangleup_{j,l}, \tau)=0
\end{equation}
The solution of eq.(C5) is
\begin{equation}
P(\bigtriangleup_{j,l}) =C(D_{\lambda, j}| \bigtriangleup_{j,l}|^2+
D_{\eta, j})^{-\overline{\lambda}_{j}/2D_{\lambda,j}-1/2}
\end{equation}
where $C$ is a normalization constant. Th PDF is then
\begin{equation}
P(\bigtriangleup_{j,l}) \propto \left \{ \begin{array}{ll}
                    {\rm const} & \mbox{$0<\bigtriangleup_{j,l}| \ll s $ } \\
                    (\bigtriangleup_{j,l})^{-\beta} &
                      \mbox{$|\bigtriangleup_{j,l}| \gg s$} \\
                   \end{array}
                \right.
\end{equation}
where
\begin{equation}
s = \frac{D_{\eta, j}}{D_{\lambda, j}},
\end{equation}
and
\begin{equation}
\beta = \frac{\overline{\lambda}_j }{D_{\lambda, j}}+1.
\end{equation}

\end{document}